\pdfoutput=1
\documentclass[12pt,
				a4paper,
				twoside,
				appendixprefix,
				notitlepage,
				BCOR8mm]{scrartcl}

\usepackage{color}
\usepackage{pdfpages}
\usepackage{pdflscape}
\usepackage{graphicx}
\usepackage{tabularx}
\usepackage{booktabs} 
\usepackage{authblk}

\begin{document}
\thispagestyle{empty}
\pagenumbering{gobble}

\makeatletter
\let\@fnsymbol\@roman
\makeatother


\titlehead{

\vspace{1cm}


}




\title{\Large Laser Wire Scanner \\
Compton Scattering Techniques \\  for the Measurement of the \\ Transverse Beam Size of Particle Beams \\ at Future Linear Colliders}




\author{\normalsize The Laser Based Beam Diagnostics Group 2001 to 2006 \\ 
I. Agapov, K. Balewski, G. A. Blair, J. Bosser, H. H. Braun, E. Bravin, \\ G. Boorman, S.T. Boogert,  J. Carter,  E. D'amico, N. Delerue, D.F. Howell, \\ S. Doebert, C. Driouichim J. Frisch, K. Honkavaara,  S. Hutchins, T. Kamps\thanks{kamps@helmholtz-berlin.de}, \\  T. Lefevre,  H. Lewin, T. Paris, F. Poirier, M. T. Price, R. Maccaferi, S. Malton,\\  G. Penn, I. N. Ross, M, Ross, H. Schlarb, P. Schmueser, S. Schreiber, D. Sertore,\\  N. Walker, M. Wendt, K. Wittenburg}

	
\date{\small 09.12.2014
}










\maketitle
\begin{abstract}
This archive summarizes a working paper and conference proceedings related to laser wire scanner development for the Future Linear Collider (FLC) in the years 2001 to 2006. In particular the design, setup and data taking for the laser wire experiments at PETRA II and CT2 are described. The material is focused on the activities undertaken by Royal Holloway University of London (RHUL).
\end{abstract} 
%
\clearpage


\clearpage
This page is intentionally left blank.



\section*{Supplementary material}

# Compton Scattering Techniques for the Measurement of the Transverse Beam Size of Particle Beams at Future Linear Collider

G.A. Blair[a], G. Boorman[a], J. Frisch[b], K. Honkavaara[c],
T. Kamps[a,*] T. Paris[a], F. Poirier[a] I. N. Ross[d],
M. Ross[b] H. Schlarb[c], P. Schmüser[c], S. Schreiber[c],
D. Sertore [e], N. Walker[c], M. Wendt[c], K. Wittenburg[c]

[a] Royal Holloway University of London, Egham, Surrey, TW20 0EX, UK

[b] Stanford Linear Accelerator Center, Stanford, CA 94309, USA

[c] Deutsches Elektron–Synchrotron DESY, D-22603 Hamburg, Germany

[d] Rutherford Appleton Laboratory, Chilton, Didcot, Oxon OX11 0QX, UK

[e] INFN Milano LASA, I-20090 Segrate (MI), Italy

December 2, 2002

Compton scattering techniques for the measurement of the transverse beam size of particle beams at future linear colliders (FLC) are proposed. At several locations of the beam delivery system (BDS) of the FLC, beam spot sizes ranging from several hundreds to a few micrometres have to be measured. This is necessary to verify beam optics and to obtain the transverse beam emittance. The large demagnification of the beam in the BDS and the high beam power puts extreme conditions on any measuring device. With conventional techniques at their operational limit in FLC scenarios, new methods for the detection of the transverse beam size have to be developed. For this laser based techniques are proposed capable of measuring high power beams with sizes in the micrometre range. In this paper general aspects and critical issues of a generic device are outlined and a specific solutions proposed. Plans to install a laser wire experiment at an accelerator test facility are presented.

*t.kamps@rhul.ac.uk

# Contents



# 1 Introduction

## 1.1 Scientific Motivation

High luminosity is the key to many of the physics processes of special interest at the Liner Collider. This fundamental point is the main physics motivation for this project and justifies considerable efforts to ensure that the machine can deliver on its excellent luminosity potential.

There are many examples of the need for high luminosity. For instance, one of the goals of the LC will be to determine the detailed nature of electroweak symmetry breaking and, if supersymmetry (SUSY) is realised in nature, to gain insight into the nature of SUSY breaking. A clear route to this goal is to measure as far as possible the entire particle spectrum and couplings with high precision and thereby to determine the structure of nature in a model independent way. As a result, particle threshold scans feature strongly in the physics LC program. They generally require high luminosity but in return provide the ultimate precision. Such an approach can produce an excellent measurement of the SUSY particle spectrum, which in turn allows the model-independent reconstruction of nature at very high energy scales (cite here worldstudy).

The case for the highest luminosities is now globally accepted and all the LC proposals currently have this as their goal, with quoted luminosities of a few$\times 10^{34}$ cm$^2$s$^{-1}$. The key motivation for this project is to add to the arsenal of tools that the machine will need to maximise its luminosity performance. In particular this project aims to provide a reliable and flexible method of obtaining real-time information on the emittance and quality of the beam and hence to allow feedback for maximising the luminosity.

## 1.2 Emittance Measurement

In this project we limit our attention to the measurement of the electron beam transverse phase space (transverse emittance) because it is the fundamental determining factor for the final transverse beam-spot size at the interaction point (IP). It is important to keep the emittance low so as to maximise the luminosity at the IP and much effort is spent in designing the accelerator and beam delivery system (BDS) to avoid sources of emittance growth. The BDS generically consists of approximately a kilometre of beam optics providing collimation, chromatic correction and final focusing. There are many potential sources of emittance growth which in general will be time dependent and will require continuous measurement and feedback to correct.

The aim is to measure the emittance of the beam to better than 10% as it approaches the IP and this will require a number of profile measurements along the BDS. In Tab. 1 beam characteristics for TESLA, CLIC and NLC are listed. Given that the beam optics are known

|  |  | CLIC | NLC/JLC | TESLA |
|---|---|---|---|---|
| BDS | $\sigma_x[\mu$m] | 3.4 to 15 | 7 to 50 | 20 to 150 |
|  | $\sigma_y[\mu$m] | 0.35 to 2.6 | 1 to 5 | 1 to 25 |
| IP | $\sigma_x^*$[nm] | 196 | 335 | 535 |
|  | $\sigma_y^*$[nm] | 4.5 | 4.5 | 5 |

Table 1: Beam spot sizes for various Linear Collider designs. Quoted are numbers for CLIC [1], NLC/JLC [2], and TESLA [3].

precisely, a set of transverse profile measurements can be translated into a determination of the emittance. At least five scanning stations will be required for each lepton beam, possibly fired by a single laser system plus laser beam transportation. Each station will need to provide a profile along three directions, as required to specify an ellipse. Relating a set of such

transverse profiles to the emittance and optimising the layout of scanning stations within a BDS design will form an interesting parallel project, that will be addressed via detailed simulations.

The electron bunch transverse profile has been measured in the past by intersecting the electron beam with a solid wire and by counting the subsequent background rate as a function of the relative position of wire and bunch. Using this technique, resolutions of typically a few $\mu$m can be obtained, at the expense of some disruption to the beam. This technique cannot be used universally at the LC, however, because the beam-spot sizes can be much smaller, the need for continuous measurement precludes an invasive technique and the intensities are so great that the wires would be quickly damaged, even if swept rapidly through the beam. For these reasons, it is necessary to develop a novel technique that can run continuously and reliably during machine operation, that does not get destroyed by the beam and that can be sufficiently fast so as to be sensitive to individual electron bunches within the bunch train. All these advantages could in principle be provided using optical scattering structures.

## 2 Laser Wire Principle

As mentioned above a laser wire measures the transverse electron beam size by scanning a narrow laser beam transversely over the electron beam. A possible setup is shown in Fig. 1. A high power laser beam is divided into two different optical paths for scanning

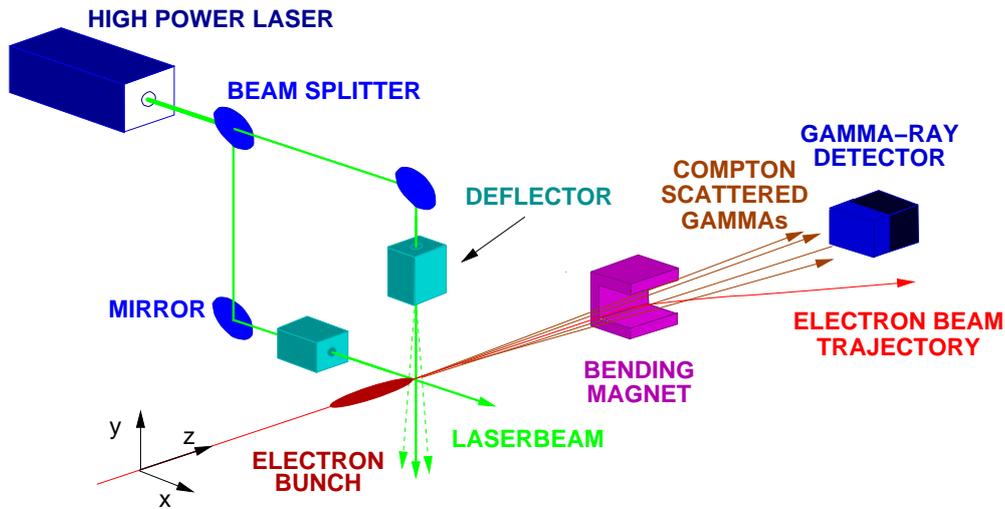

Figure 1: Operation principle of a laser wire profile monitor.

the horizontal and vertical beam size. Before the interaction with the electron beam the laser beams are focused. The electron beam is then bent away while the Compton scattered photons travel along a straight line where they are detected with a calorimeter. Scattered electrons will be bent more strongly than particles with the nominal beam energy enabling detection at a location after the bending magnet.

The proof of principle of this technique was first demonstrated at the Stanford Linear Collider (SLC) [4] and a different design is presently being tested at the Accelerator Test Facility (ATF) at KEK [5, 6, 7]. The SLC design used a high power pulsed laser beam, transported some distance to the IP within the SLD experiment. The very tight space and accessibility of this location led to a highly engineered design, with the laser position fixed in space. The operation of the device then required the electron beam to be scanned across the laser beam and, in this sense, this laser wire was still an invasive device. The ATF design uses a continuous wave laser (with corresponding low peak power) and the entire optical system is moved relative to the electron beam using stepping motors.

Our aim is to elevate these designs to a compact, non-invasive device where a high-power pulsed laser is scanned across the electron beam using either piezo-driven mirrors or acousto-optic devices. These techniques have not been used before in a laser wire and so we are proposing a genuine advance in this field.

## 3 Optical Scattering Structures

Common to all optical scattering structures is that they must have features smaller or similar in size to the particle beam under measurement. Several types of laser spot structures can be generated with common optical setups. In the following some optical structures are listed together with their performance rating.

### 3.1 Laser Wire (Gaussian Profile)

The laser beam is here focused to a small gaussian spot with radius $\omega_o$. If we consider a diffraction limited, finely focused beam waist, the minimal achievable spot radius is given by $\omega_o = \lambda/(\pi\theta)$, where $\lambda$ denotes the laser wavelength and $\theta$ the half opening angle of the laserbeam at the waist (see Fig.2). The distance over which the laser beam diverges to $\sqrt{2}$

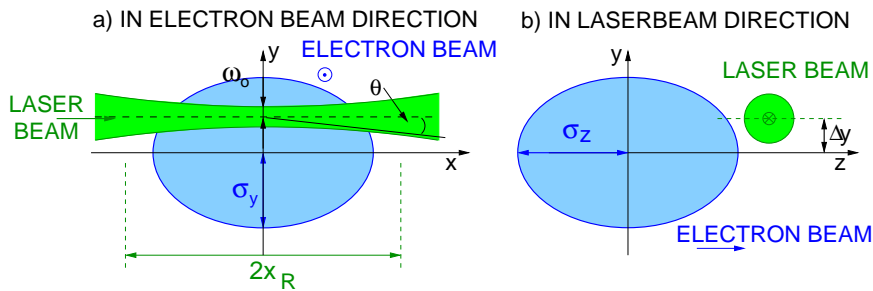

Figure 2: Scheme of a gaussian laser beam focused to its diffraction limit scanned over an electron beam.

of its minimum size is called the Rayleigh range $x_R$ and defines the usable length of the laser wire. The smallest achievable spot size with diffraction limited optics is in the order of $\omega_o \sim \lambda$ (See the section on laser optics for more details). With Nd:YLF or YAG laser working at higher harmonics electron spot sizes from $\sigma_y > 350$nm can be measured with high accuracy. The laser beam power must be in the order of a couple of MW to yield a few thousand Compton photons per scan spot. Critical issues of a laser wire design are the diffraction limited optics, which must withstand such a high beam power and the scanning system, enabling intra-train scanning of consecutive bunches.

### 3.2 Laser Wire (Dipole Mode)

The resolution of the laser wire can be enhanced by generating an artificial transverse dipole mode by means of a lambda half waveplate, where half of the gaussian is shifted in phase by $90^o$. Such a waveplate can easily be installed in the optical path of the laser wire and would enhance the resolution of the device by roughly a factor of two aiming at beam sizes in a region from 250nm $< \sigma_y < 500$nm.

### 3.3 Laser Interferometer

Towards beam sizes in the nanometer range, a standing wave interference pattern generated by crossing two laser beams has been proposed and successfully tested at the FFTB experiment [8].

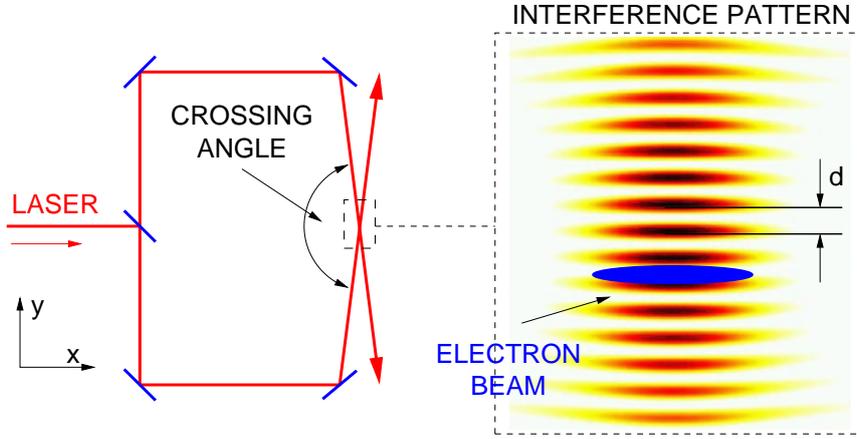

Figure 3: Scheme for the generation of an interference pattern with fringe spacing $d$.

The fringe spacing of the interference pattern (see Fig. 3) depends on the laser wavelength and on the crossing angle. The electron beam is moved over the pattern and the Compton scattered photons are detected. If the beam size is small compared to the fringe spacing, a modulation of the Compton signal is observed which is proportional to the transverse electron beam size. This modulation vanishes if the beam size is large compared to the fringe spacing. The smallest observed spot size with this technique was about 58 nm [9].

## 4 Physics Processes

### 4.1 Compton Scattering

We collect here some Compton scattering results which are relevant to our analysis. The relativistic Compton process has been studied in detail elsewhere [10]. Relevant aspects for our analysis have been presented in Ref. [9]. The Compton cross section $\sigma_C$ is related to the Thomson scattering cross section $\sigma_0 = 6.65 \times 10^{-25} \text{cm}^2$ by:

$$\sigma_C = \sigma_0 \frac{3}{4} \left\{ \frac{1+\epsilon}{\epsilon^3} \left[ \frac{2\epsilon(1+\epsilon)}{1+2\epsilon} - \ln(1+2\epsilon) \right] + \frac{1}{2\epsilon} \ln(1+2\epsilon) - \frac{1+3\epsilon}{(1+2\epsilon)^2} \right\} \tag{1}$$

where $\epsilon = \gamma \frac{\omega_0}{m_e}$ is the normalised energy of the laser photons in the electron rest frame and $\gamma$ is the Lorentz factor associated with the incident electron beam energy $E_b$ ($\gamma = \frac{E_b}{m_e}$). The energy spectrum of the emerging gamma rays is given by.

$$\frac{d\sigma}{d\omega} = \frac{3\sigma_0}{8\epsilon} \left\{ \frac{1}{1-\omega} + 1 - \omega + \left[ \frac{\omega}{\epsilon(1-\omega)} \right]^2 - \frac{2\omega}{\epsilon(1-\omega)} \right\} \tag{2}$$

with maximum final scattered photon energy of $E_{\max} = 2E_b\epsilon/(1+2\epsilon)$. The angular distribution of the scattered photons is sharply peaked in the direction of the electron motion and confined in a cone with opening angle a few times the critical angle $\alpha_c$, which is defined with

$$\alpha_c = \frac{\sqrt{1+2\epsilon}}{\gamma}. \tag{3}$$

The Compton cross section $\sigma_C$ is in Fig. 4 evaluated for two scenarios: One for a typical

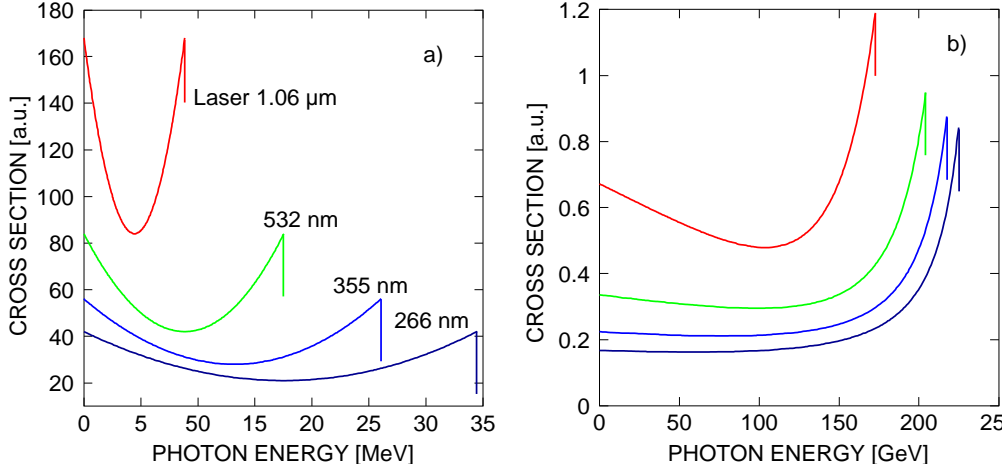

Figure 4: Compton cross section for the first four harmonics of an Nd:YAG laser scanning a 1 GeV (a) and 250 GeV (b) electron beam.

linear collider test facility beam energy (1 GeV), where sub-systems of a LWS will be tested and for a typical linear collider beam delivery system energy (250 GeV). Currently a Nd:YAG based laser system is the instrument of choice because of its performance capabilities with respect to high power and small spot size.

Now that the entire electron bunch sees the laser beam, the number of Compton photons per electron bunch is given by:

$$N_\gamma = N_b \frac{P_L \sigma_C \lambda}{c^2 h} \frac{1}{\sqrt{2\pi}\sigma_s} \exp\left(\frac{-y^2}{2\sigma_s^2}\right) \qquad (4)$$

where $N_b$ is the number of electrons per bunch, $\sigma_s^2 \equiv \sigma_y^2 + \sigma_o^2$ the overlap region (Note that the laser beam radius is defined as $\omega_o = 2\sigma_o$), and $y$ is the transverse position of the laser spot. See Fig. 2 for this definitions.

We can now use these results in Eqn. 4 to calculate the peak number of Compton photons per bunch, $N_C^{\text{bunch}}$ which occurs when the offset $y=0$

$$N_C^{\text{bunch}} = N_b \frac{P_L \sigma_C \lambda}{c^2 h} \frac{1}{\sqrt{2\pi}\sigma_s}. \qquad (5)$$

### 4.1.1 Approximations

Applying the above assumptions together with a bunch occupancy of $6\times10^9$ (1 nC) and beam energy of 1 GeV, gives

$$N_C^{\text{bunch}} = 303 \left(\frac{P_L}{10\text{MW}}\right) \left(\frac{\lambda}{350\text{nm}}\right) \left(\frac{30\mu\text{m}}{\sigma_s}\right) \left(\frac{N_b}{6\times10^9}\right) \qquad (6)$$

It should be noted that the Compton cross section is at low electron beam energies only a tiny correction to the Thompson cross section, which is energy independent.. The maximum final scattered photon energy is $E_{\text{max}} = 2E_b\epsilon/(1+2\epsilon)$. Parameters relevant to us give to good approximation

$$E_{\text{max}} = (1+\delta) \cdot 6.71 \text{ MeV} \qquad (7)$$

$$\sigma_C = (1 - 0.0136\delta) \cdot 6.56 \times 10^{-25} \text{cm}^2 \qquad (8)$$

where

$$\delta \equiv \left[\left(\frac{E_b}{500 \text{ MeV}}\right) \left(\frac{350 \text{ nm}}{\lambda}\right) - 1\right] \text{ MeV} \qquad (9)$$

Alternativly

$$E_{\max} \propto E_b^2/\lambda \qquad (10)$$

so

$$E_{\max} \sim \left(\frac{E_b}{500\text{MeV}}\right)^2 \left(\frac{350\text{nm}}{\lambda}\right) 6.7\text{MeV}. \qquad (11)$$

## 4.2 Background Processes

Backgrounds are all processes which lead to an additional energy deposit in the detector and hence underlay the Compton signal. These backgrounds must be reduced to the extent that the signal to background ratio can be optimised. This background study is of particular importance as it can lead to specific experimental strategies eg to build background shielding or increase the laser power.

Simulations are currently carried out using the GEANT3 [11] FORTRAN based toolkit which covers electromagnetic interactions down to 1 keV. The physics processes handled by the toolkit models include photoelectric effects, Compton scattering, bremsstrahlung and ionisation. Simulations were also performed within the GEANT4 [12] C++ based framework with a low energy electromagnetic package [13] to extend the energy simulation down to 250 eV. No major differences were found. However, GEANT3 is found to be more stable than GEANT4 (Version 4.3.0) and so GEANT3 is chosen for the simulations presented below.

Specifically Synchrotron Radiation, beam gas bremsstrahlung, and thermal photons signals are studied. These three processes are the main backgrounds which produce photons in the PETRA environment, where we plan to install a laser wire experiment.

### 4.2.1 Synchrotron Radiation

Synchrotron radiation (SR) occurs when a high energy charged particle enters a magnetic field and its trajectory is bent. A subsequent loss of energy is observed as an emission of photons. The critical photon energy $E_c$ is often used to define SR as it marks the spectral point for which one-half of the total power is irradiated at lower photon energies, and one-half at higher (see Fig. 5). For a bending magnet with angle $\theta_{dip}$ and length $L_{dip}$ the critical energy is given by [14]

$$E_c = \frac{3\hbar}{2} \cdot \frac{c\gamma^3\theta_{dip}}{L_{dip}} \qquad (12)$$

with $\gamma = E_b/m_e$ the Lorentz factor of the particle beam. For the background simulations the generation of the SR is performed with a Monte Carlo algorithm described in [15] which makes use of the critical energy and the spectral distribution.

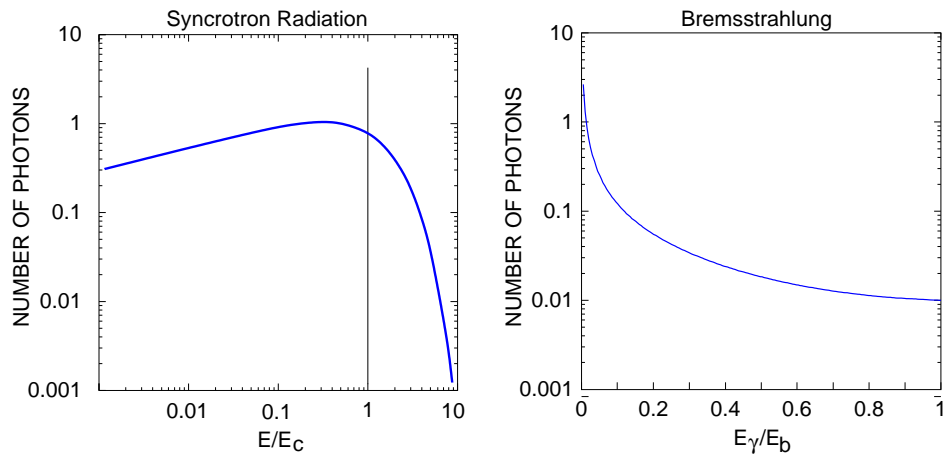

Figure 5: Photon spectrum for synchrotron radiation and bremsstrahlung.

The number of SR photons $N_{\gamma_{SR}}$ which are on average emitted per particle in a bunch, seen by a detector of acceptance width $A$ as illustrated in Fig.6, can be computed using

$$N_{\gamma_{SR}} = \frac{5\alpha}{2\sqrt{3}} \gamma \frac{dL}{R} \tag{13}$$

with $R$ the radius of the dipole magnet, $dL$ the beam length over which photons emitted tangential to the beam will enter the detector acceptance width $A$ and $\alpha = 1/137$ the fine structure constant.

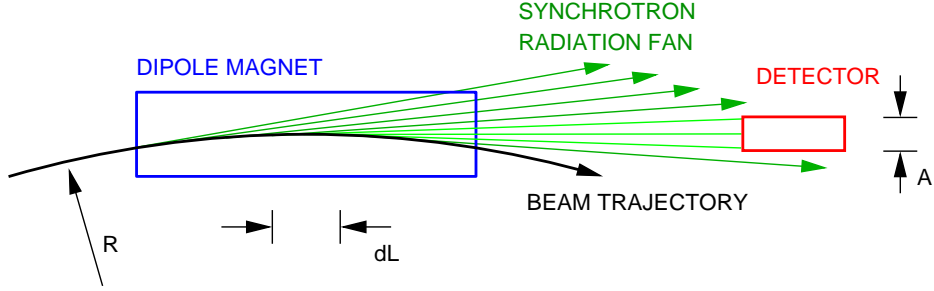

Figure 6: Beam trajectory in dipole magnet with SR fan and detector.

### 4.2.2 Bremsstrahlung Scattering

Besides SR the beam can loose particles via interactions with the residual gas in the beam pipe region. These interactions are elastic and inelastic scattering. At high energies the bremsstrahlung process is dominant, where beam electrons emit photons while they are deflected by the electric field within the gas atom. The resulting gamma-rays produce a continuous spectrum from zero to the energy $E_b$ of the incoming positron as plotted in Fig. 5.

### 4.2.3 Thermal Photons

A third background component is produced by thermal photons radiation. Thermal photons originate from the blackbody radiation of the beampipe gas. The radiated photons interact with the positron beam via the Compton process and then give rise to scattered photons. The average energy of the radiated photons $E_{\gamma_{th}}$ is:

$$E_{\gamma_{th}} = 2.7 k_B T \simeq 2.3 \cdot 10^{-4} [T] eV \tag{14}$$

where $k_B$ is the Boltzmann constant and $T$ is the absolute temperature of the blackbody in Kelvin. At room temperature ($T \simeq 300 K$) the average photon energy is approximately 0.07 eV. The density $p_{\gamma_{th}}$ expressed in number of photons per cubic meter within the beampipe is given by

$$p_{\gamma_{th}} = \frac{2.4(k_B T)^3}{\pi^2 (c\hbar)^3} \simeq 20.2 \cdot T^3 \tag{15}$$

which for room temperature gives $p_{\gamma_{th}} = 5.33 \cdot 10^{14}/m^3$. The rate of such collisions per bunch can then be calculated using

$$n_{\gamma_{th}} = p_{\gamma_{th}} L_{bmp} N_e \sigma_T \tag{16}$$

with $N_e$ the number of electrons (positrons) per bunch, $L_{bmp}$ the length of the beampipe and $\sigma_T$ the Thomson cross section.

**Thousand Photons: LC and LCTF**

| Parameter | | TESLA | CLIC | NLC | TTF2 | PETRA | ATF | CTF2 Drive | CTF2 Probe | CTF3 Drive |
|---|---|---|---|---|---|---|---|---|---|---|
| Beam energy | [GeV] | 250 | 500 | 250 | 1.0 | 4.5/7/12 | 1.28 | 0.04 | 0.045 | 0.150 |
| Electrons/bunch | [$10^9$] | 20 | 20 | 20 | 6 | 75 - | 6 | 10 - 600 | 3.75 | 14.5 |
| Bunches/pulse | | 2820 | 154 | 95 | 2820 | - | 1 | | | |
| Bunch spacing | [ns] | 337 | 0.667 | 2.8 | 337 | 480/192/96 | 2.8 | | | |
| Macro pulse length | [$\mu$s] | 950 | 0.140 | 0.266 | 950 | | | | | |
| Repetition rate | [Hz] | 5 | 100 | 120 | 5 | - | - | 5-10 | 5-10 | 10 |
| Hor. bunch size | [$\mu$m] | 20 | 3.4 | 7 | 55 | 300 - 200 | 50/10 | 200 | | |
| Ver. bunch size | [$\mu$m] | 2 | 0.34 | 2 | 55 | 30 - 20 | 5/1 | 200 | | |

# 5   Expected Signals

Using the equations in Sec. 4.1 the expected scattered photon flux $N_C$ and maximum scattered photon energy $E_{max}$ for the different machine options can be computed. Results are listed in Tab. 2 for a laser peak power of $P_L = 10$ MW and laser spot size of $\omega_o = 1$ $\mu$m with $\lambda = 532$ nm.

|            | TTF2 | PETRA7 | PETRA4.5 | ATF  | CTF2  | CTF3  |
|------------|------|--------|----------|------|-------|-------|
| $N_C$      | 254  | 352    | 1093     | 2726 | 158   | 228   |
| $E_{max}$ [MeV] | 17.5 | 777.5  | 334.6    | 28.6 | 0.003 | 0.514 |

Table 2: Compton photon flux and maximum photon energy for various machine options. The number of Compton photons scales with $N_C \propto P_L \cdot \lambda \cdot (\omega_o^2 + \sigma_y^2)^{-1/2}$.

# 6 Laser System

## 6.1 Requirements

The laser wire beam profile monitor puts strigent requirements on the laser system to be used. The requirements for any kind of laser include excellent fundamental mode quality and diffraction limited performance. The wavelength of the laser should be in the order of $\lambda \sim 500$ nm (PETRA and TESLA) or even in the UV with $\lambda \sim 250$ nm (TESLA) for smaller spot sizes and thus higher resolution. The laser peak power should be in the order of $P_{max} = 3$ MW (TESLA) to 1 MW (PETRA) in order to obtain 1000 photons per single pulse interaction. If a scanning scheme with $n$ pulses per scan spot is desired, the required laser power reduces to $P = P_{max}/n$. Ideally the laser pulse structure is locked to the pulse structure of the electron beam. The pulse structure for TESLA and PETRA is depicted in Fig. 7.

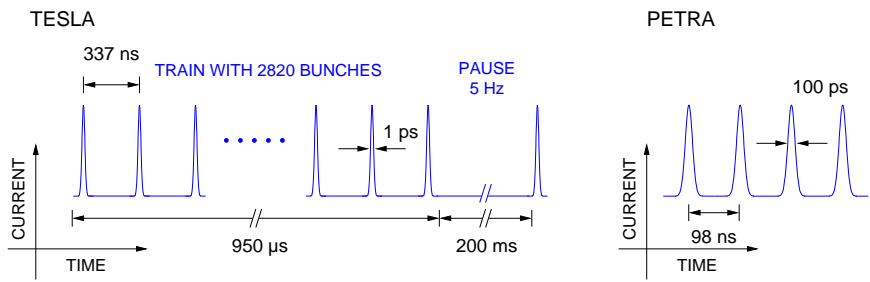

Figure 7: TESLA and PETRA macro pulse structure.

## 6.2 Options

### 6.2.1 TESLA

Because of the low duty cycle of TESLA ($\eta = 0.01$) a pulsed laser with the same macro pulse structure seems most apporiate. Roughly 3000 bunches are in one train of which 15k appear per second. In order to measure across one bunch train a laser with a high repition rate is required. This could be a laser system like used for the TTF photo injector (see Fig. 8). Such a laser system is installed at the TTF and delivers $E = 100$ $\mu$J energy per pulse at $\lambda = 523$ nm with $\Delta t = 10$ ps pulse length. The laser is a mode-locked Nd:YLF laser with several harmonic generators delivering light in the UV. The repition rate of the laser is $f_{rep} = 5 - 10$ Hz. Since the laser is uesed to illuminate a photo cathode its beam profile is flat and not gaussian. In order to facilitate such a laser some more R&D work is required to obtain the required pure gaussian transverse profile. Along this some work is necessary to reach a compact design of the laser system allowing installation in the accelerator tunnel in minimum space.

### 6.2.2 PETRA

The high duty cycle of PETRA (repition rate $f_{rep} = 130$ kHz) could possibly allow the use of a cw laser with low average power. Allowing for one second scan time to integrate the incoming signal requires in the order of $P = 10$ W laser power. The costs of a 10 W cw laser unit are in the same order as for pulsed laser system. A pulsed system would consist of a Q-switched laser with 10 ns pulse length and 100 mJ energy per pulse. Due to the longitudinal pulse structure of Q-switch pulses (spiking) an injection seed is needed. Such lasers are widely used in industry and commercially available from many companies..

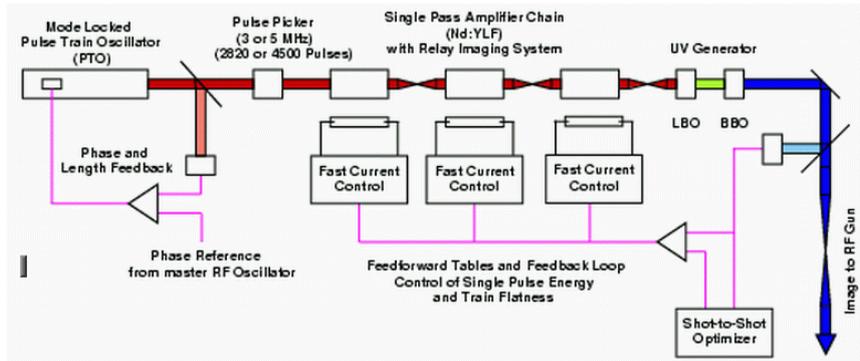

Figure 8: TTF photo injector laser system.

### 6.2.3 Conclusion

For TESLA a multi-pulse mode-locked laser sytem is necessary to enable profile measurements during in one bunch train. For PETRA it is envisaged to start experiments with a single pulse laser optimizing the signal to noise ratio. To ease trigger and to obtain stability the cw option should be considered as an extension.

## 7 Laser Scanning

- Mechanic; linear or rotating system
- Acousto-Optic
- Combination of both
- Other methods
- Slow or fast scan

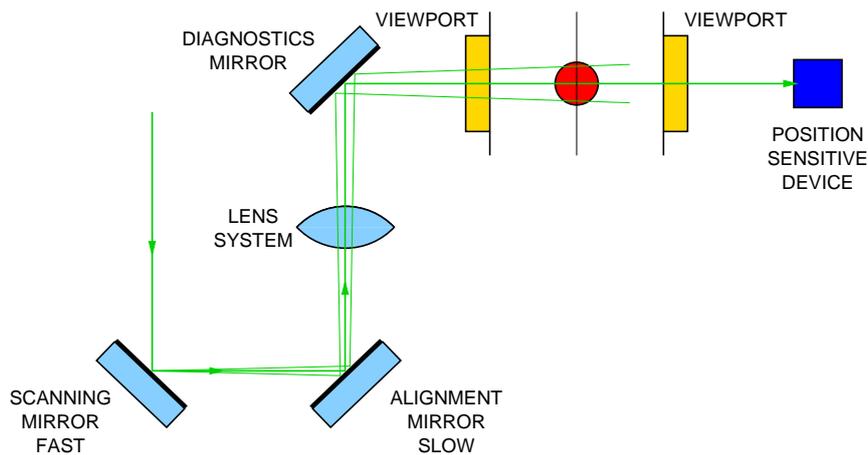

Figure 9: Schematic layout of a scanning system.

## 8 Detector and DAQ

### 8.1 Compton photon detector

The requirements for a photon calorimeter are listed below:

- Short decay time $< 1000$ ns to avoid pile up effects in the material (1MHz bunch frequency).

- Radiation hardness to survive operation in synchrotron radiation environment.

- High resolution and high light output to obtain a good signal to noise ratio.

- Small Moliere radius to obtain a small shower in transverse dimension to limit the required space for the detector.

The energy deposit in the detector from the Compton process can be estimated by taking the total number of photons multiplied with the mean energy of the photon spectrum. For the PETRA case with $4.5$ GeV electron beam energy this results in $E_{dep} = 183$ GeV. This energy is confined in a cone with opening angle a few times the critical angle of the Compton photons which is for the above case $\alpha_c = 0.12$ mrad. On top of that energy from background processes such as multiple gas scattering and synchrotron radiation has also to be taken in account.

Among the materials meeting these requirements are scintillating glass (Ce+) and lead tungsten crystals (PbWO). Lead tungsten crystals are favourable because of positive results from Compton polarimeter experiments at CEBAF [16]. The individual crystals of the detector matrix are read out by standard photomultiplier tubes supplied with positive high voltage.

## 8.2  Scattered Electron Detector

## 8.3  Data Acquisition

# 9 Laser Optics

## 9.1 Basics

One important aspect of the Laser wire project is the laser optics. The focusing optics must be able to focus the laser beam to appropriate spot size $\omega_o$ and with confocal parameter $b = 2z_R$ for using it as a laser wire. For the focusing system, an aberration corrected optics design must be developed to focus the intense laser beam inside the electron beam chamber. Special care must be taken in order to avoid damage on any optical components caused by high beam power.

### 9.1.1 Laser spot size

The laser beam is focused to a small spot size $\omega_o$. There are a number of factors that affect spot size. The most important are the diffraction limit, the transverse mode quality $M^2$, and the spherical aberration. Two contributions to the spot size are defined. One contribution comes from paraxial optics, which is the gaussian spot size. The other contribution is due to aberrations, mainly spherical, which is present as a function of lens shape, orientation and index of refraction.

From accelerator considerations one supposes an electron bunch with a gaussian distribution of particles in the plane transverse to the bunch velocity. Before starting with spot sizes, let us see how a gaussian beam is defined and what is exactly the waist of a beam. The term Gaussian refers to the illumination intensity of the beam. The intensity is symmetric around the beam axis and has a gaussian shape described by

$$I(r) = \frac{2P}{\pi\omega} \cdot \exp\left(-\frac{2r^2}{\omega^2}\right) \tag{17}$$

where $r$ is the radial distance and $\omega$ the radius of the beam at the $1/e^2$ point of the intensity, commonly called the beam spot size or $1/e^2$ criterion. The beam radius is related to the RMS spot size by $\omega = 2\sigma$. The spot size contains approximately 86% of the total power in a gaussian beam. If we want the spot size to contain 99% of the total power we need to go to $r = \pi\omega$.

### 9.1.2 Gaussian optics

When results of paraxial optics (on-axis) are applied off-axis, the theory is known as Gaussian optics. In this case, a gaussian shaped laser beam is focused to a small gaussian spot size radius $\omega_o$. The minimum achievable spot radius is given by the diffraction limited beam waist ($TEM_{00}$ or $M^2 = 1$ mode) [17]:

$$\omega_o = \frac{\lambda}{\pi\theta} \tag{18}$$

where $\lambda$ is the wavelength of the laser and $\theta$ the divergence or half opening angle of the laser beam at the waist. If the beam travels in $x$-direction, the propagation of the waist is described by

$$\omega(x) = \omega_o \cdot \sqrt{1 + \left(\frac{\lambda x}{\pi\omega_o^2}\right)^2} = \omega_o \cdot \sqrt{1 + \left(\frac{\theta x}{\omega_o}\right)^2} \tag{19}$$

Another important parameter to be taken into account is the distance a collimated gaussian laser beam can propagate without spreading significantly. This is important because we need the laser to be collimated in a region comparatively longer than the electron beam dimension parallel to the laser beam. The distance travelled by the beam before the beam diameter increases by $\sqrt{2}$ is called the Rayleigh range $x_R$ and is defined with

$$x_R = \frac{\omega_o}{\theta} = \frac{\pi\omega_o^2}{\lambda} \tag{20}$$

The Gaussian laser beam is an idealisation which is never reached in practise. A real laser can be well described by a diffraction limited Gaussian beam, and a quality factor $M^2$ is defined as the relative beam size and divergence with respect to the diffraction limited Gaussian beam. Therefore, the quality factor $M^2$ enters Eq. 18 and we obtain the gaussian spot size for a laser mode $M^2$

$$\omega_o = \frac{M^2 \lambda}{\pi \theta} \qquad (21)$$

In terms of requirements for beam size measurements, let us consider an horizontal scan of the electron beam, as shown in Fig. 2. In this case $\sigma_y$ must be aligned perpendicular to the direction of the laser beam. A transverse laser beam size at the interaction point less than half the vertical dimension of the electron beam, $\omega_0 < \sigma_y$ and a Rayleigh range greater or equal than the horizontal dimension of the electron beam $x_R \geq \sigma_x$ are necessary to reach a good resolution in using the laser beam as a spot monitor. An equivalent argument is valid for the electron beam transverse dimension $\sigma_x$.

Since the embedded laser radius is defined in the $1/e^2$ criteria, when a gaussian laser beam of radius $\omega_{in}$ is focused by a lens of focal length $f$, the divergence $\theta$ of the beam can be well approximated in thin lens theory by $\theta \simeq \omega_{in}/f$. The $f$-number $f\#$ of a focusing lens is defined as the lens focal length divided by the filled diameter at the lens, $f\# = f/D$. Since we are talking about the filled diameter at the lens, here it is more convenient to use the 99% criteria, that is $D = \pi \omega_{in}$. Using this definition the diffraction limited spot size, Eq. 21, can be set as a function of the $f$-number:

$$\omega_o \approx M^2 \lambda f\# \qquad (22)$$

Therefore, the smaller the $f$-number the smaller the spot size $\omega_o$. Diffraction is an inescapable result of the wave nature of light; it places a limit on the performance of an optical system. Diffraction causes light beams to spread transversely as they propagate. The most important thing to note about diffraction is that it increases linearly with focal length but is inversely proportional to the beam diameter.

### 9.1.3 Real optics

If a perfect lens is used to focus a collimated beam, the spot size is limited only by diffraction. But a real optical system can not be simulated using only paraxial optics. One has to take into account lens shape. In terms of spot size, the most important aberration is the so called spherical aberration. This kind of aberration occur when, tracing rays far from the axis, the image rays converge closer to the lens than the gaussian focal point. This situation is shown in Fig. 10. Spherical aberration has the effect of increasing the spot size as well as causing the best focus to occur at different location than the calculated focal length. Spherical aberration is a function of lens shape, lens orientation and lens index of refraction.

The exact spot size for a given lens and circumstances must be determined by raytracing, but there is a useful formula for determining an estimate of spot size due to spherical aberration

$$\omega_{sph} = \frac{kD^3}{2f^2} = \frac{kD}{2f\#^2} \qquad (23)$$

where $k$ is a function of the shape of the lens, the index of refraction of the lens and the wavelength of the laser. In this formula the spot size due to the spherical aberration is proportional to the cube of the beam diameter and inversely proportional to the square of the focal length.

The diffraction contribution to spot size is independent of lens shape, while the aberration contribution is dependent on lens shape through the parameter $k$. Therefore it is mainly when the aberration contribution becomes important that lens shape becomes important, and that happens at low $f$-numbers. The spot size contribution due to diffraction (Eq. 22) increases with $f\#$ while the spherical aberration spot size (Eq. 23) decreases with $f\#$. Adding up the two contributions may not be very rigorous, but it gives a conservative estimation for

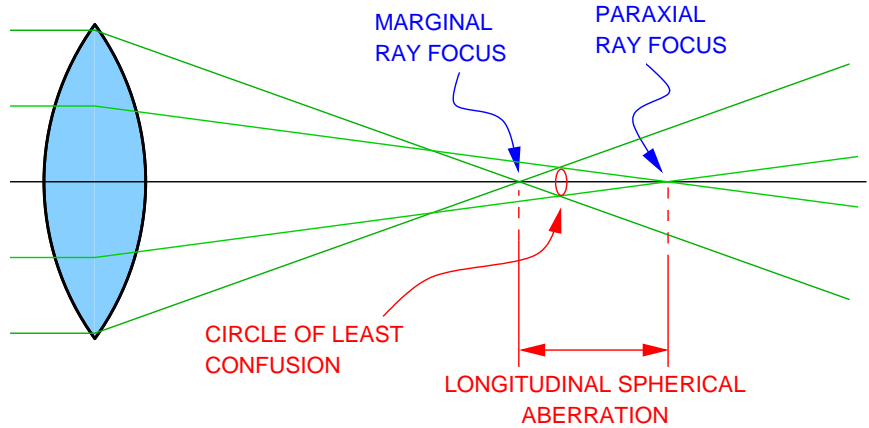

Figure 10: Spherical aberration has a longitudinal and transverse component. The longitudinal spherical aberration reduces the focal point respect to the paraxial ray focus. The transverse spherical aberration produces an increase in the spot size. The circle of least confusion is the point where the maximum concentration of rays is found in the minimal spot size; that is the real focus.

the real spot size. Therefore the minimum achievable spot size for a given lens is obtained through a compromise between the focal length $f$ and the diameter of the input beam $D$.

## 9.2 Requirements for PETRA and CTF2 experiments

The focus optics should enable basic scanning and measurement of the transverse electron beam size at the CTF2 with reasonable resolution. To achieve a minimum resolution in the order of half the beamsize under measurement $\sigma_y$ the laser focus should not exceed the electron beam size $\sigma_o \leq \sigma_y$. The electron beam is assumed to be round with dimension $\sigma_{x,y} \simeq 200$ $\mu$m. This leads to a spot size of $\sigma_o = 5 - 10$ $\mu$m. Even with a poor quality of the fundamental mode ($M^2 \leq 20$) the laser spot size is in the order of the electron spot size.

The focal length of the optics should be BFL $\geq 150$ mm in order to incorporate a folding mirror for diagnostics in the path between optics and best focus. .

In addition the optics system should be easy to align, be durable, contain a minimum number of elements, and be cost effective.

### 9.2.1 CTF2

The scanning will be provided by stepping motors on which the whole optics system (scan mirror, lens and diagnostics mirror) is mounted. Subsequently there is no need for a telecentric design. Furthermore the system must be able to withstand the high peak power of the proposed laser (100 MW at 10 ps pulselength with $\lambda = 1.064$ $\mu$m)

### 9.2.2 PETRA

For the PETRA a different scanning system is planned. Options are a acousto optic or a piezo driven system. Depending of the size of the scanning device, the scanner can be mounted either before or after the focusing lens. The laser power at PETRA will be much lower than at CTF2, since only 2 MW peak power are necessary to obtain 1000 photons per interaction. Currently a $ns$ pulsed laser with less than 10 MW at $\lambda = 532$ nm is projected.

## 9.3 Options

Various lens system have the capabilities to conform with the requirements listed above. Under investigation were single lenses, telescopes, cemented and air-spaced achromats, and

off axis parabolic mirrors.

Single lenses and telescopes were rejected because these techniques cannot provide diffraction limited performance with the required spot size. Cemented achromats are not usable in our case because of damage threshold issues. The damage threshold for an average cemented achromat is $750\,\mathrm{MW/cm^2}$ for cw laser and $0.75\,\mathrm{J/cm^2}$ at $10$ ps for pulsed lasers.

### 9.4 Simulations and Results

Simulations using the ray tracing code ZEMAX were performed in order to evaluate different focus optics options. The problem with the code is that it ignores interference and diffraction. Ray tracing is based on light propagating along straight rays through homogeneous media. ZEMAX determines the geometrical spot size as the minimum RMS spot radius. In addition, ZEMAX can calculate the propagation of an embedded gaussian beam using the diffraction limited techniques mentioned in the previous section. Since we are interested in the spot size produced by an optical system, we can make good estimation of it by adding up the gaussian spot size (gaussian calculation) and twice the RMS spot radius (geometrical calculation) in quadrature

$$\omega_o^2 = \sqrt{\omega_{gauss}^2 + (2\sigma_{rms})^2} \tag{24}$$

#### 9.4.1 Achromats

The lens systems under investigation are listed in Tab. 3 and the setup used for the simulations is sketched in Fig. 11. All systems perform within 3% variation in a common manner. In Fig. 12 the results for the Linos lens system is plotted.

The achievable spot size scales linearly with $M^2$. Even with $M^2 = 20$ the spot size is $\sigma_o = 200\,\mu\mathrm{m}$ (Note: $\omega_o = 2\sigma_o$). With a $2.5$ mm input beam radius this results in a Rayleigh Range of $z_R = 800\,\mu\mathrm{m}$, which is sufficient for the experiment. While all lenses show almost the same performance optically, the damage threshold differs due to different coatings and lens material. The CVI lens damage threshold is already not sufficient, leaving the Linos and OptoSigma lens as two alternatives. Ghosting analysis to second order bounces has been performed with all lenses. With the setup as depicted in Fig. 11 no ghost spot have been found inside the lenses. However, a full simulation including all optical elements from the spatial filter on are necessary to identify possible sources for higher order spots.

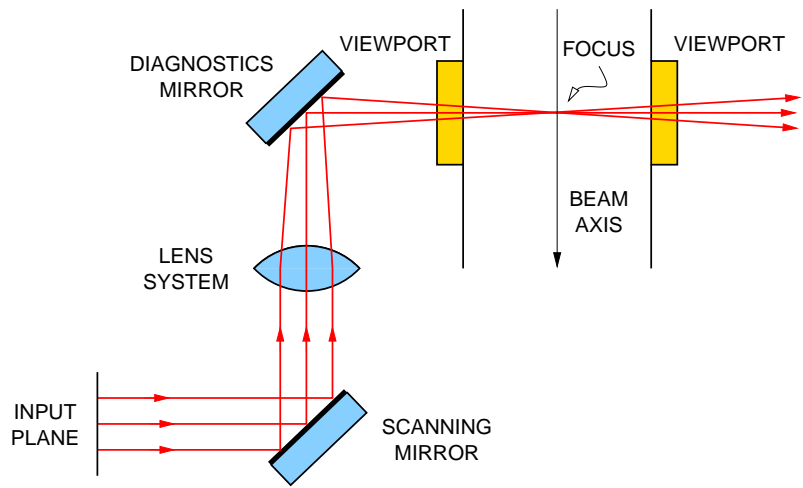

Figure 11: Setup used for simulations with ZEMAX. The lens system defines the space for the system under investigation.

The quoted air spaced achromats can handle up to $10\,\mathrm{J/cm^2}$ for $10$ ns long pulses. For a pulse with a given energy the damage threshold scales with the inverse square root of

| Provider | Part | BFL [mm] | Material | Damage [J/cm$^2$] | Price |
|---|---|---|---|---|---|
| OptoSigma | 027-0180 | 143.8 | ? | 8 | 830.00 EUR |
| Linos | 03 8910 | 151.99 | K5, SF2, and SF11 | 10 | 535.00 EUR |
| CVI | LAP 150 30 | 141.70 | SF11 | 5 | 783.50 EUR |

Table 3: Air spaced achromats under investigation. All systems are available with anti-reflective coating for IR light and in stock at the supplier. Damage threshold numbers are given for 10 ns long pulses.

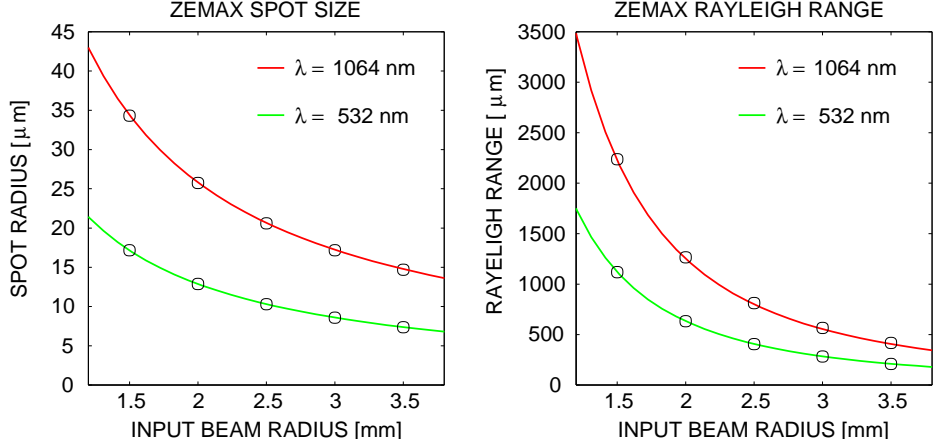

Figure 12: Results from ZEMAX simulations with the Linos achromat at $\lambda = 1.064~\mu$m and $\lambda = 532$ nm. Circles show simulated points, the line is a fit using spot size equations. The calculated spot radius is $\omega_o$ at $1/e^2$.

the pulse length. This relation is deduced from the equation of heat transfer and has been proved for pulses with length above 1 ns. For pulses below 1 ns the damage threshold should be constant because the pulse is much faster than the heat transfer so that no heat can be transported out of the irradiated area during the pulse (Source: Linos). If we scale the pulse length from 10 ns down to the given 10 ps by the inverse square root relation, we should have a worst case estimation. We get a value of $0.316$ J/cm$^2$ as a worst case estimation. The laser energy coming from the spatial filter is 1 mJ with M$^2 = 1$. For a 2.5 mm beam radius this gives $0.040$ J/cm$^2$ (using $\sigma = \omega/2$), a factor of eight lower than the damage threshold.

### 9.4.2 Parabolic Mirrors

Off axis parabolic mirrors provide true diffraction limited performance and a higher damage threshold. The main drawback of these mirrors is the high sensitivity to alignment. In the following some mirrors from Janos Technology are investigated against their use at the CTF2 laser wire laser focus. See Tab. 4 for details on these and Fig. 13 for results concerning their diffraction limited performance.

Mirrors can cope with beam power up to GW/cm$^2$ giving a good argument to use a parabolic mirror.

It seems that off axis parabolic mirrors may be useful for the laser focus since their performance is highly compatible with the air spaced achromats. A possible setup to use mirrors at the CTF2 laser wire is sketched in Fig. 14

### 9.5 Discussion

Both systems, air spaced achromats as well as parabolic mirrors, offer performance characteristics within the requirements for the laser wire experiments at CTF2 and PETRA. However,

| Part | Diameter [mm] | f Parent [mm] | Effective FL [mm] | Price [USD] |
|------|---------------|---------------|-------------------|-------------|
| A8037-262 | 50.80 | 38.10 | 76.20 | 332.00 |
| A8037-202 | 50.80 | 50.80 | 101.60 | 332.00 |
| A8037-206 | 50.80 | 76.20 | 152.40 | 332.00 |

Table 4: Characteristics of some parabolic mirrors from Janos Technology. All mirrors are made of aluminium with gold coating. The reflectivity for IR light is 99%. The spot size can be calculated using the effective focal length.

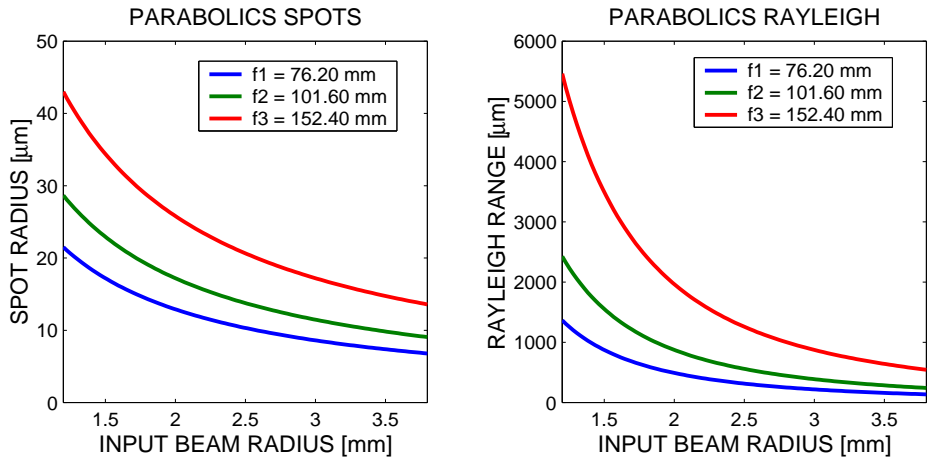

Figure 13: Performance of off axis parabolic mirrors with three different focal length.

there are differences between the systems. While the achromats are easier to operate, the alignment of the parabolic mirrors demands more labour. Simulations with ZEMAX show that the spot size at the focal length increases for tilts in elements of the optical line. Allowing a tilt of $\pm 0.5^o$ increases the spot size by 1% for the achromats and by a factor of three for the parabolic mirror. Tilting by $\pm 1^o$ increases by 2% for the achromats and by a factor of six for the parabolic mirror. From the damage threshold point of view, all options are usable. It should be mentioned that the mirror requires gold coating since the standard metal surface is not suitable for high power lasers. The most cost effective focus elements is the Linos achromat followed by the mirrors leaving the other two achromats, from OptoSigma and CVI, the most expensive alternative. Combining all these arguments indicate that the Linos air spaced achromat provides the most efficient focus.

### 9.6 Next Steps

After decision on which system will be used at both experiments is necessary to verify the performance of the system, including the viewports, using the test bench at RHUL. Furthermore more detailed information about the incoming laser beam characteristics (above all $M^2$) should be provided to enable better estimates on the system performance.

## 10 Laser Optics Measurements

Spot size measurements with the proposed lens system (Linos HALO) were done in order to study the focusing, beam propagation around the best focus, and tolerances. Measurements include spot size scans with and without a viewport window. The aim was to achieve an error of less than 2 % on the local spot size enabling comparision with ZEMAX simulations.

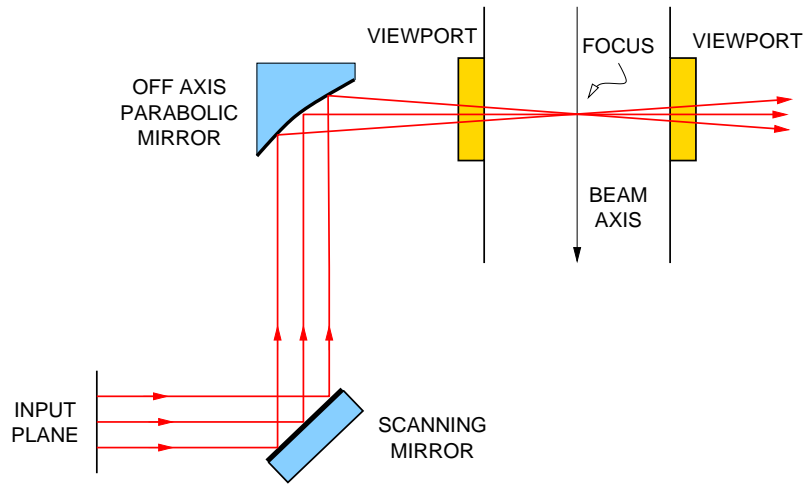

Figure 14: Setup using parabolic mirror to focus the incoming laser beam into the beam chamber.

## 10.1 Strategy

Knife edge scans with a piezo driven razor blade were chosen as measurement technique because of the simplicity of the measurement and the high precision.

Among other techniques are pinhole scans or slit scans. These methods are unable to measure spot sizes in the sub-$\mu$m range with the required precision and do not conform with ISO standards. With the pinhole method and small aperture is moved between the laser beam and a large-area photo detector. The photo current versus the distance is the profile. A general rule of thumb is that the aperture should be between 1 % and 30 % of the smallest spot to be measured. Real world decisions often skew the decision to values approaching 30 %. With the knife edge technique a sharp edge is passed between the beam and a photo detector, the knife edge scan (photo current versus distance) trace must differentiated. This can either happen electronically or numerically on the data. Another possibility is to fit directly the integral over the beam distribution to the data in order to obtain the beam size. The last possibility has the disadvantage that one assumes already a specific kind of distribution such as gaussian. But a pinhole reveals the true intensity distribution for any beam, gaussian or not. The drawback of the pinhole is that it must be accurately positioned in two dimensions relative to the beam; achieving this is often time-consuming. The disadvantage of a knife edge is the large range of powers reaching the detector in the course of a single measurement. Both methods suffer from diffraction effects from the aperture. These have to be counteracted, for example with a focusing lens.

Since the beam spots under investigation are in the $\mu$m range the knife edge techniques was chosen.

## 10.2 Measurement Setup and Procedure

In Fig. 15 the measurement setup is sketched. The laser light coming from the green HeNe laser[1] is first guided with a mirror to a collimator. The collimator is a Keplerian telescope consisting of two plano-convex lenses[2] with a magnification factor of almost six. After magnification the enlarged laser beam is bend by another mirror before arriving at a flat beam splitter, where half the power is guided into a Michelson interferometer while the other half passes another mirror before going through the the focusing lens. The beam power going into the interferometer arm is split up again with a cubic beam splitter. One part of the light

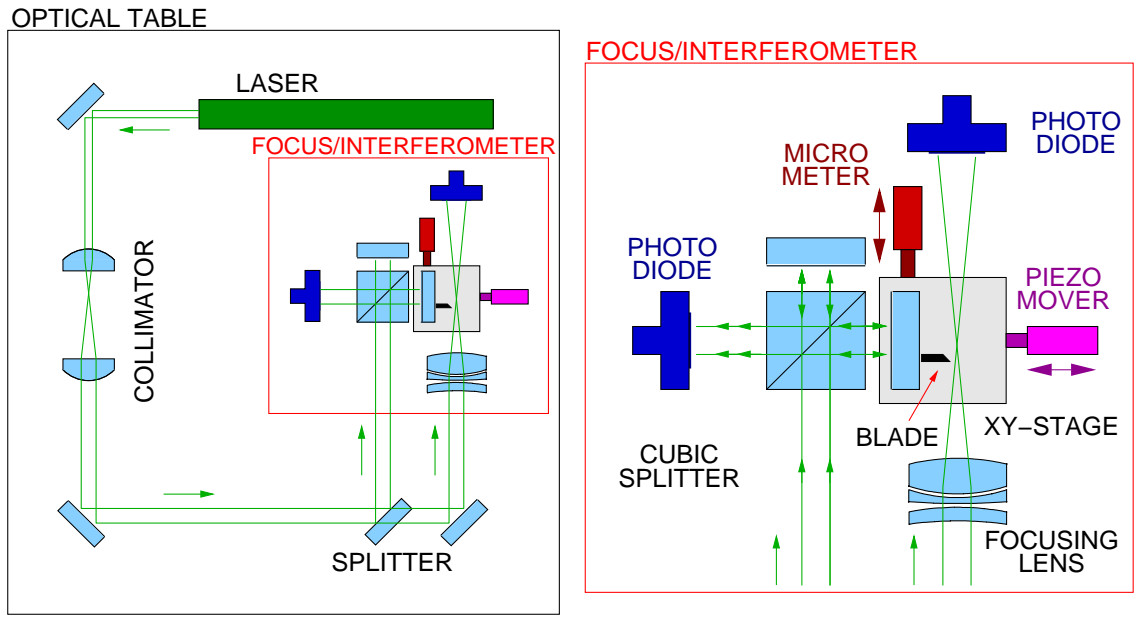

Figure 15: Measurement setup used for spot size scans. On the left the complete overview and on the right the scan and interferometer unit zoomed.

is reflected by the mirror with the knife edge mounted on before it is recombined again with the other half and detected by a photo detector. The photo current is read out by signal processing electronics and converted into a voltage signal. Following the the focusing lens the focused beam is collected by another photo detector with similar readout. The knife edge together with its mirror is moved by a piezo electric driven screw in forward and reverse direction through the laser beam as sketched in Fig.16. The knife edge is a shaving razor blade

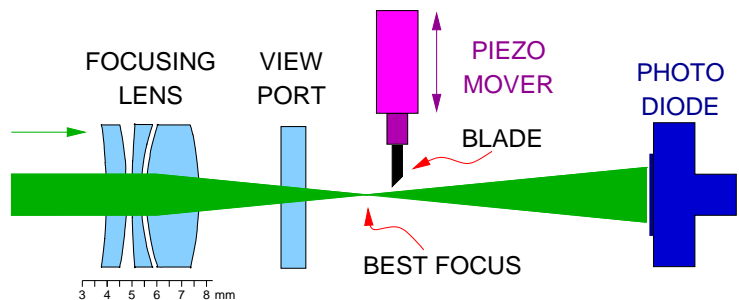

Figure 16: Knife edge section of the measurement setup.

with an edge roughness of less than 0.2 nm. The two photodetector signals are fed into an ADC card which is plugged into a PC running a LabView application, which also controls the movement of the piezo driven screw. With the interferometer readout it is possible to calibrate the movement of the piezo driven screw with very high precision. The average step width if the piezo driven screw is in the order of 20 nm depending on the load of the screw and hysteresis effects.

For one complete scan the beam profile was measured in forward and reverse direction at eleven longitudinal positions of the focusing lens. This was repeated with a viewport window in the beam path between focusing lens and knife edge.

### 10.2.1 Interferometer

Beforehand each measurement cycle the interferometer arm had to adjusted to get the center of the interference pattern centered on the photo detector. An aperture was then placed in front of the detector to enhance the contrast of the light shining on the photo diode area. The top plot of Fig. 17 shows normalized data for a fraction of the path the piezo driven screw was moved while scanning one laser beam slice. On the average 10000 points were scanned for one slice. The distance between two minima or two maxima is $\lambda/2$. The detected voltage

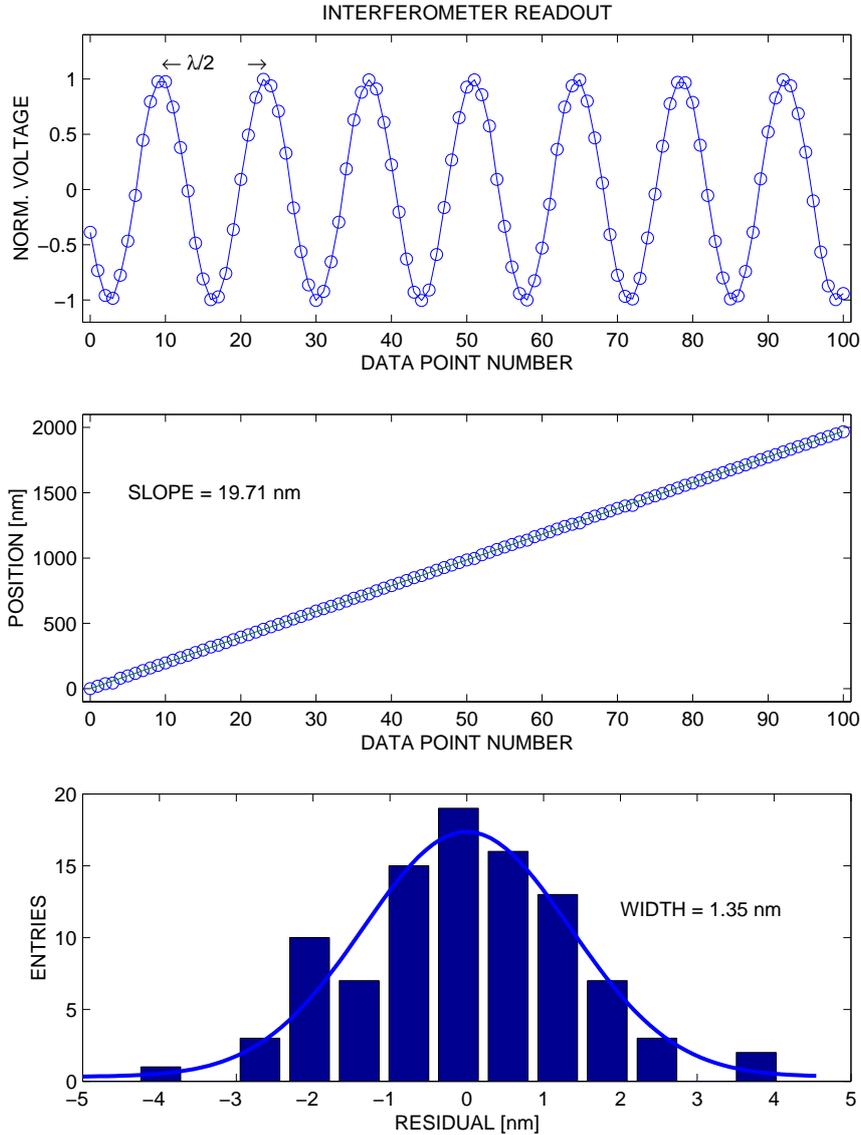

Figure 17: Interferometer raw data and anlysis plots. Top: Normalized data from the photo detector. Middle: Voltage values converted into position moved by piezo driven screw plus straight line fit. Bottom: Residual plot of position moved per step.

while moving the mirror with the knife by distance $d$ is

$$V_{DI} = V_{oI} \left( 1 + \cos \left( \frac{4\pi d}{\lambda} + \pi \right) \right) \tag{25}$$

so the distance $\Delta d$ the piezo driven screw with the razor blade moved each step is then

$$\Delta d = \frac{\lambda}{4\pi} \arccos \left( \frac{\Delta V_{DI}}{V_{oI}} \right) \tag{26}$$

The middle plot of Fig. 17 shows the periodically changing voltage $V_{DI}$ converted into distance covered $d$ versus the data point number. A straight line fit is applied to extract the average step width of the piezo driven razor blade. For analysis of the knife edge photo detector data the position information was used directly not the average step width.

### 10.2.2 Knife Edge Scan

In Fig. 18 the output of the photo detector on the knife edge side is depicted. The data reflects

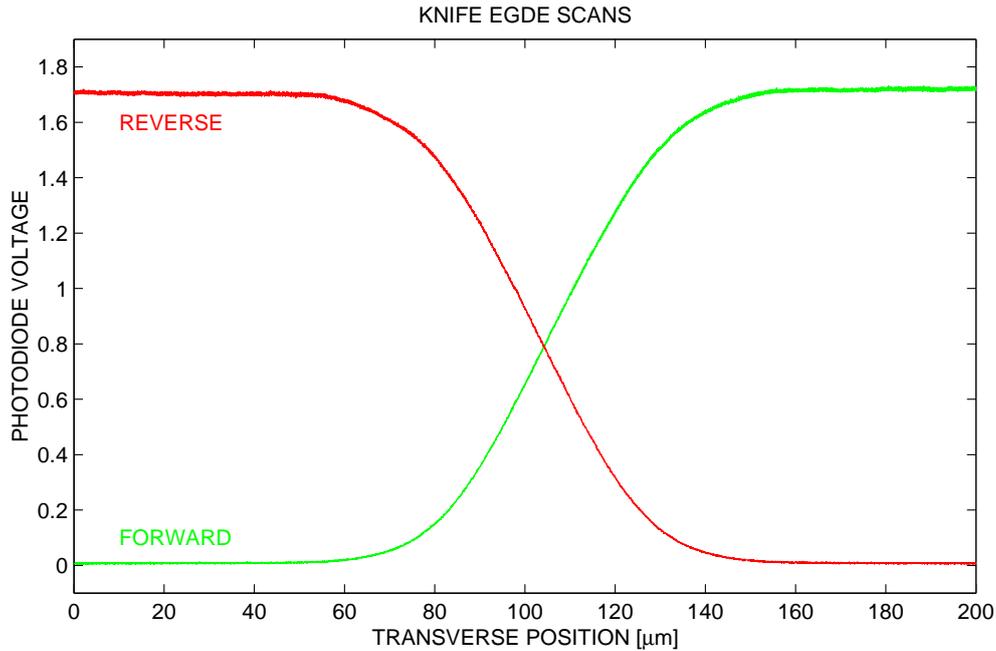

Figure 18: Photo detector output versus transverse position for one laser beam slice.

the integral of the the transverse laser beam distribution. Assuming a gaussian distribution the voltage data $V_{DK}$ can be fitted using the errorfunction according to

$$V_{DK} = V_{oK} + V_{1K} \cdot erf\left(\frac{x - x_{off}}{\sqrt{2}\sigma}\right) \qquad (27)$$

where $x$ the transverse position, $x_{off}$ the transverse offset from center position, and $\sigma$ the laser beam waist RMS radius. Such plots as in Fig. 18 were obtained for eleven different longitudinal position along the laser beam focus. Furthermore data was taken with an without viewport window. The propagation for the these two cases is depicted in Fig. 19 The focus for the scenario with viewport window is shifted towards the photo detector.

### 10.2.3 Results

The propagation of the beam envelope $\sigma(z)$ along the longitudinal axis $z$ is described by

$$\sigma_z = \sigma_o \left(1 + \left(\frac{M^2\lambda(z - z_{off})}{4\pi\sigma_o^2}\right)^2\right)^{1/2} \qquad (28)$$

where $\sigma_o$ describes the minimum beam size. This function was fitted to the data obtained from the complete scans with and without viewport. Free parameters for the fit were the longitudinal offset $z_{off}$ and the minimum beam size $\sigma_o$. The mode quality is $M^2 = 1.03$ for the laser used in our experiment. This can be calculated from the inital beam paramters at the laser head exit and are $d_o = 0.88$ mm for the beam diameter at the $1/e^2$ points and $\theta =$

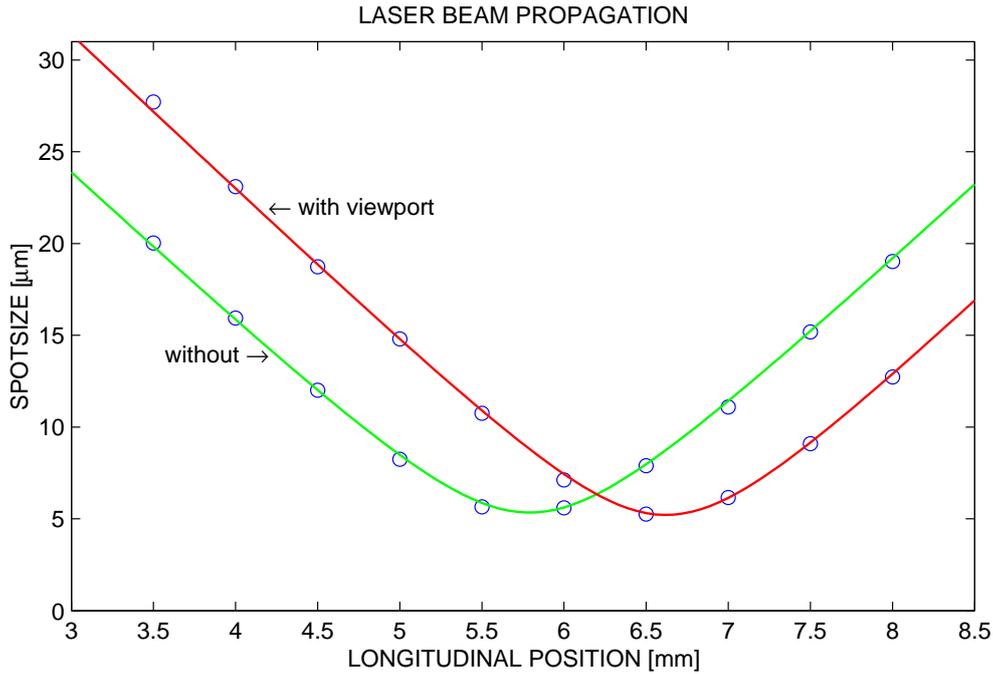

Figure 19: Propagation of the beam envelope with and without the viewport window.

|                  | $\sigma_o$ [$\mu$m]              | $z_{off}$ [mm]                  |
| ---------------- | -------------------------------- | -------------------------------- |
| with viewport    | $5.34 \pm 0.02 \pm 0.07$         | $5.79 \pm 0.01 \pm 0.08$         |
| without viewport | $5.21 \pm 0.03 \pm 0.07$         | $6.62 \pm 0.01 \pm 0.08$         |

Table 5: Results for the minimum beam size and longitudinal offset with and without viewport window. The values are grouped with their statistical and systematical error.

0.81 mrad for the beam divergence. Results for the free parameters are listed in Tab. 5 For the minimum spot size the main contributor to the systematical error are diffraction effects from the razor blade edge. These will be further investigated in Sec. 10.3. The systematical error of the longitudinal offset is dominated by misreading of the mechanical micrometer screw the lens system is attached to. Surprinsingly the spot size seems to shrink a little bit when the viewport is added to the beam path as one would expect that the viewport material would defocus the beam. Within error margins the two spots (with and without) overlap. It should be mentioned that the spot slices without viewport window were measured in forward and reverse direction while the spot slices with the viewport windows were measured only in forward direction. At the start of the reverse measurement the piezo driven screw broke down due to a technical failure. Some more investigation is necessary to verify this effect.

The focal spot is moved by $\Delta z_{off} = 0.83 \pm 0.01 \pm 0.08$ mm when the viewport is inserted. As a rule of thumb any spot moves by one third of the thickness of any flat inserted into a focused beam. Since the viewport has a diameter of $t = 2.50$ mm this value agrees nicely.

### 10.2.4 Input Beam

The input beam radius was measured by removing the focusing lens out of the beam path. It was measured to have a radius of $\sigma_{in} = 1.227 \pm 0.02$ mm, where the error reflects the statistical error.

## 10.3  Diffraction Effects

In order to calculate the intensity measured in the photodiode as a function of position of the razor, we need to include any diffractive effects. Referring to Fig. 20, $x$ is the position of the razor with respect to the optical axis of the system, where $L$ is the distance between razor and photodiode, which in our case is $L = 12.5$ cm. The wavelength of the laser light used is

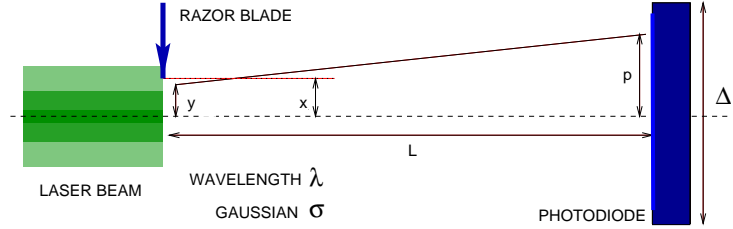

Figure 20:  Schematic of the layout of the beam waist measurement illustrating the terms used to calculate Fresnel effects.

$\lambda = 532$ nm. The intensity $I$ of the laser light is given by:

$$I = I_0 \exp(-\frac{y^2}{2\sigma^2}) \tag{29}$$

where $\sigma$ is the gaussian width of the beam and $y$ is the distance of the point of measurement from the optical axis. The value of $\sigma$ depends on the position of the lens and varies, for the laser wire measurements, from a minimum of approximately 5 $\mu$m at the laser waist, to about 1 mm for the input parallel laser beam.

The amplitude $A$ of the electromagnetic wave at a point on the photodiode that is a distance $p$ from the optical axis is given by:

$$A(y,p) = A_0 \exp(-\frac{y^2}{4\sigma^2}) \exp(i\frac{2\pi}{\lambda}r) \tag{30}$$

where

$$r = \sqrt{L^2 + (p-y)^2} \tag{31}$$

Integrating Eq. 30 over the aperture defined by the position of the razor gives $A_{\text{tot}}(x,p) = A_{\text{Re}}(x,p) + iA_{\text{Im}}(x,p)$ where

$$A_{\text{Re}}(x,p) = A_0 \int_{-\infty}^{x} e^{-\frac{y^2}{4\sigma^2}} \cos\left[\frac{\pi}{\lambda L}(y-p)^2\right] dy \tag{32}$$

$$A_{\text{Im}}(x,p) = A_0 \int_{-\infty}^{x} e^{-\frac{y^2}{4\sigma^2}} \sin\left[\frac{\pi}{\lambda L}(y-p)^2\right] dy \tag{33}$$

The total intensity at $p$ is :

$$I(x,p) \propto |A_{\text{tot}}(x,p)|^2 \tag{34}$$

The results for two cases were evaluated numerically. Firstly the case of a large beamspot size $\sigma = 1$ mm, shown in Fig. 21.

Here the effects of Fresnel diffraction are clearly apparent, reflecting the change in phase seen at a point on the photodiode across the 1 mm-size wavefront at the razor. However, although the intensity fluctuates locally, the integrated power at the photodiode is not changed. This was checked by evaluating numerically the resulting photodiode output $V_{PD}$, given by:

$$V(x)_{PD} \propto \int_{-\frac{\Delta}{2}}^{\frac{\Delta}{2}} I(x,p) dp \tag{35}$$

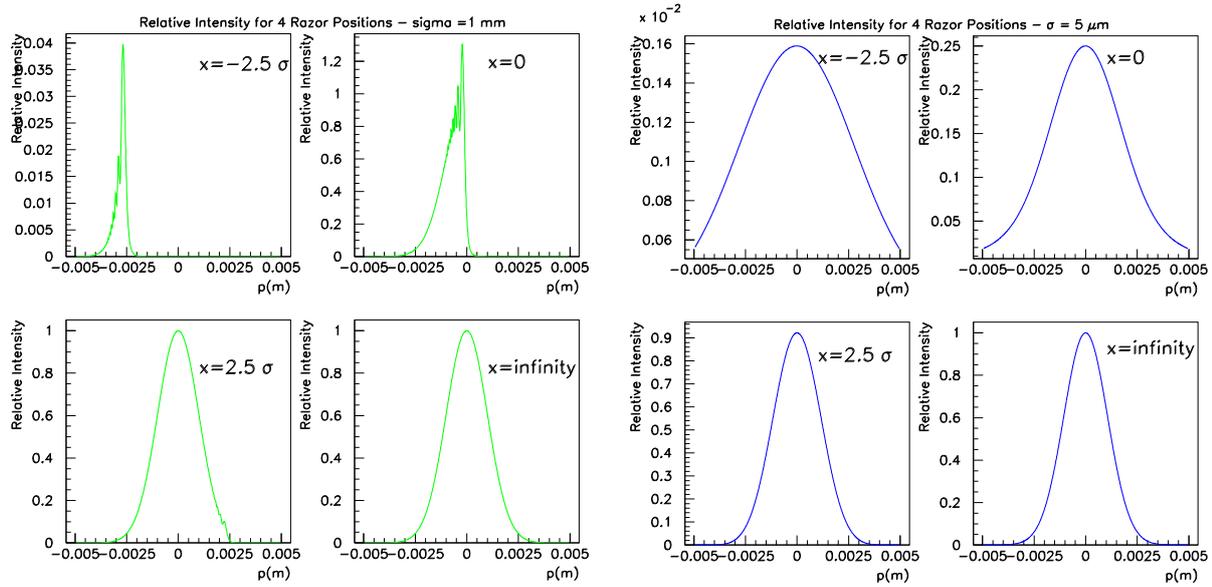

Figure 21: Left: Intensity as a function of transverse position along the photodiode for a beamspot size of 1 mm. The effects of Fresnel diffraction are clearly visible and must be integrated over the surface of the photodiode. Right: Intensity as a function of transverse position along the photodiode for a beamspot size of 5 $\mu$m. The effects of diffraction are effectively to expand the gaussian width at the surface of the photodiode .

Turning to the case of a 5 $\mu$m value of $\sigma$, the situation is different. In this case the spot size is so small that there is negligible phase variation across the spot at the razor as seen by a point on the photodiode. In this limit, the diffractive effect is to produce a gaussian intensity distribution at the photodiode, with a greater width corresponding to the smaller effective aperture set by the razor position. The integral of Eq. 34 was evaluated numerically and the results are shown in Fig. 21.

In the case of an infinite sized photodiode, the increased effective width of the gaussian would not matter. However the photodiode used was circular with total area 100 mm$^2$, corresponding to a radius of $\sim$5.6 mm. The integral of Eqn. 35 was evaluated numerically assuming a square photodiode with various values of $\Delta$, to give the resultant intensity as a function of $x$, corresponding theoretically to the measured photodiode output. By fitting this function to that expected for a gaussian beam profile, the expected $\sigma_{\mathrm{measured}}$ is obtained for each value of $\Delta$. By comparing this to the true value of $\sigma_{\mathrm{true}}$=5 $\mu$m, an estimate of the systematic error due to diffraction is obtained. The results are shown in Fig. 22 as a function of $\Delta$, the photodiode transverse dimension. Given that the diameter of the photodiode used is $\sim$ 1.2 cm, it is clear that the measurement is becoming sensitive to diffractive effects, where the effect could be of order 0.04 $\mu$m. Allowing for some conservatism and for the assumption used above of a square photodiode, this study suggests that a systematic error of order 0.05 $\mu$m should be attributed to diffractive effects for the smaller spot sizes.

## 10.4   Comparison with Simulations

The simulation with ZEMAX included the full beam length, the beam expander, and the focusing lens. For input beam the paramters from the manufacturer were chosen. This results in an input beam radius of $\sigma_{inZ} = 1.305$ mm against a measured value of $\sigma_{in} = 1.227 \pm 0.02$ mm. For a beam expander build with a telescope it is important to have the correct distance between the two lenses. The measurement setup allows only a very coarse

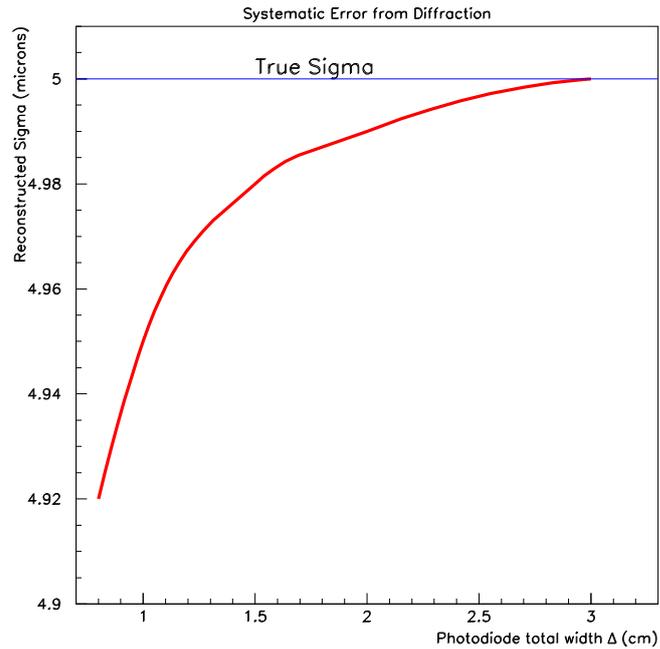

Figure 22: The effect of varying the effective size of the photo detector, assuming a square detector of width $\Delta$ .

longitudinal alignment of these elements which may lead to this discrepancy. Another effect could arise from the laser output devlivering a beam not quite according to specifications. Anyway with the given setup and under the assumption of an ideal spacing of all elements a minimum spotsize of $\sigma_{oZ} = 5$ $\mu$m can be achieved. In the next step the measured input beam radius was chosen as inital beam paramter for the simulation, resulting in a minimum spot size of $\sigma_{oZ} = 5.35$ $\mu$m. The effect of the viewport window was found to be negligable. The longitudinal shift of the minimum spot was simulated as change in the back focal length $\Delta BFL = 0.79$ mm (measured $\Delta z = 0.83 \pm 0.08$ mm).

## 11 Experiment at PETRA

In this section information on the on-going efforts for a laser wire experiment at the PETRA (Positron Electron Tandem Ring Accelerator) machine at DESY is gathered.

### 11.1 Experimental Layout

PETRA is a storage ring for electron and positrons, serving as pre-accelerator for the HEAR complex. It is planned to upgrade the machine in the near future enabling linear collider related damping rind studies. In its current state the accelerator is able to store electrons or positrons at 4.5, 7, and 12 GeV. with a beam lifetime of at least 10 hours. It can also provide a minimum bunch separation time between 480 and 192 ns with a bunch charge of the order of $10^{10}$ particles per bunch (see Sec. 5 for a full list of PETRA parameters), parameters similar to those foreseen for TESLA.

In May 2001 preliminary background measurements were performed at 4.5 and 7 GeV in two different locations (see Fig.23). These locations are foreseen to be two options for the future detector of the laser wire experiment. Location 1 (LOC1) is based at the beginning of a straight section behind a dipole whereas location 2 (LOC2) is situated approximately 98 m away at the end of the straight section. The detector in LOC2 is positioned behind a dipole and placed between a short vertical magnet and a quadrupole. The pressure inside the beampipe is measured using seven pressure monitors lying along the straight section.

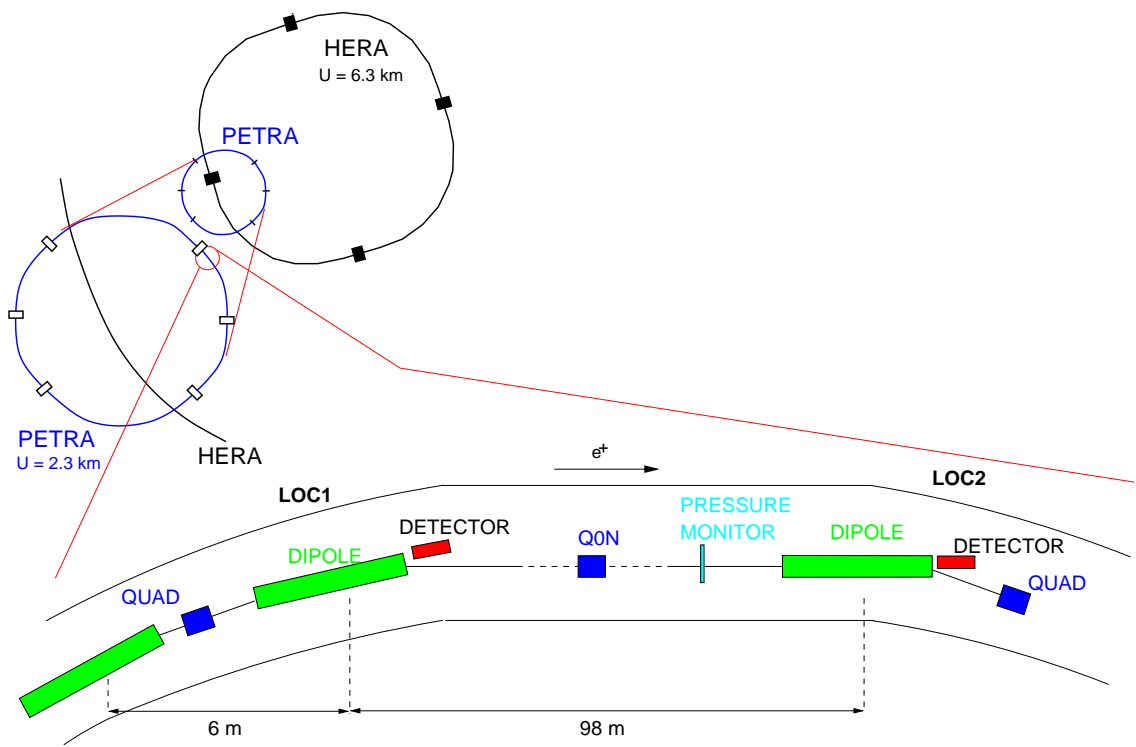

Figure 23: Schematics of the PETRA accelerator within the DESY complex and zoom into the straight section with the two detector locations.

### 11.2 First Background Simulations

A close experimental setup is displayed in Fig.24. The photons produced enter the aluminium wall of the beampipe and are then detected. The simulation setup includes a vacuum beampipe, a dipole, an aluminium beampipe wall and a detector as shown in Fig.24.

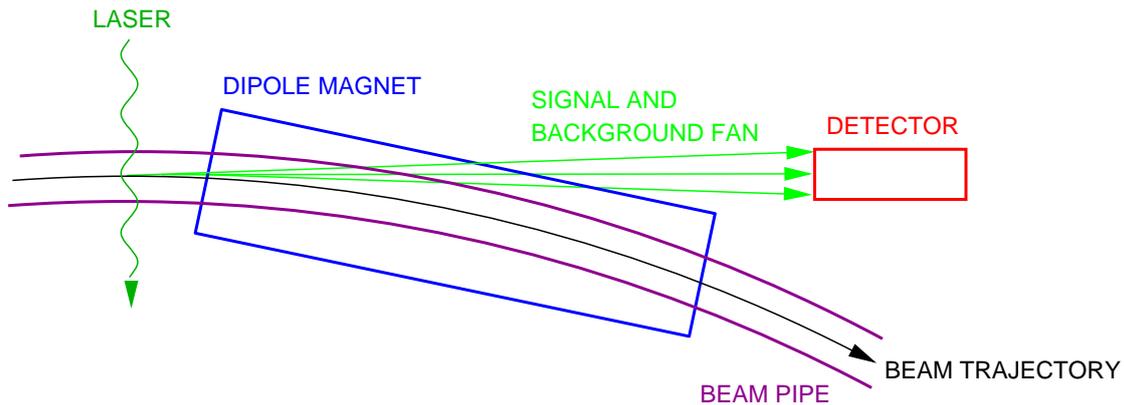

CLOSE EXPERIMENTAL SETUP

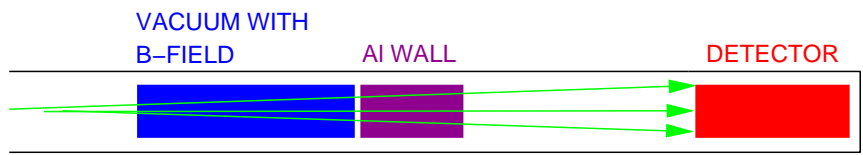

SIMULATION SETUP

Figure 24: Close experimental and simulation setup.

### 11.2.1 Synchrotron Radiation

Synchrotron Radiation backgrounds were simulated using the methods described in Sec. 4.2.1 for two PETRA electron beam energies. The critical photon energies are $1$ keV (for $4.5$ GeV) and $4$ keV (for $7$ GeV) and the number of emitted photons per positron is $0.231$ (for $4.5$ GeV) and $0.360$ (for $7$ GeV). The contribution of SR photons to the overall background spectrum is shown in Fig. 25.

### 11.2.2 Bremsstrahlung

Simulation over 130000 bunches (corresponding to $1$ sec measurement time) is performed with GEANT3 including a straight section of $10$ m length for LOC1 and $98$ m for LOC2. The vacuum pressure inside the beampipe is set to $1.5 \cdot 10^{-9}$ mbar. The contribution of bremsstrahlung photons is shown in Fig. 25.

### 11.2.3 Thermal Photons

The Compton photon energy contribution of thermal photons is shown in Fig. 25 with a maximum energy of approximately $10.8$ MeV and $26.2$ MeV for positron beam energies of $4.5$ GeV and $7$ GeV respectively.

### 11.2.4 Simulation Conclusions

Backgrounds have to be added in such a way that the contribution from two or more processes is allowed in one event. This is performed using an algorithm calculating the resulting energy $\langle E_{tot} \rangle$ from two or more processes, as described below.

The background processes are considered independent. For two backgrounds $a$ and $b$, $\langle E_{tot} \rangle$ is calculated using the following equation

$$\langle E_{tot} \rangle = \int_{E_{tot}=0}^{\infty} E_{tot} f(E_{tot}) dE_{tot} = \int_{E_{tot}=0}^{\infty} \int_{x=0}^{E_{tot}} f_a(E-x) f_b(x) dx \, dE \qquad (36)$$

where $f_a$ and $f_b$ are respectively the probability density function of backgrounds $a$ and $b$, $E_{tot}$ is the total energy of the resulting event (energy from process $a$ plus energy from process $b$). A third process can be added by considering $a + b$ as an other single process. Results of the various background simulations are depicted in Fig.25 for an energy range between 0 and 600 keV. The results have pointed out that the expected number of events detected $\mu$ follows a Poisson distribution because of the low number of photons reaching the detector. Thus, for each background processes the probability $P_n$ for n photons detected after a single bunch is given by

$$P_n = \frac{\mu^n}{n} e^{-\mu} \qquad (37)$$

Results from SR simulation have shown that a negligible number of photons go through the beampipe material at $4.5$ GeV (upper limit $\approx 1 \cdot 10^{-15}$ photons per bunch), but at a beam energy of 7 GeV peaks are observed. Due to Poisson statistics a high peak is obtained when no photons are observed, and following peaks arise for cases when one, two or more photons reach the detector. Beam gas bremsstrahlung results show that photon spectrum covers the whole range of energy studied.

## 11.3 First Background Measurements

### 11.3.1 Detector Setup and Calibration

Background measurements were carried out at both locations using a CsI(TI) crystal of size $15 \times 15 \times 100$ mm³ glued to a photomultiplier Hamamatsu R268. The light-tight box, as shown in Fig.26, containing the crystal and the photomultiplier was positioned tangential to the beampipe and 30 cm away from the dipole magnets to avoid any electromagnetic effects on the detector. As depicted in Fig. 27, the signal from the photomultiplier is sent via a charge sensitive pre-amplifier to a shaping amplifier. The bipolar output of the shaping amplifier is used to generate a clock signal via a discriminator and a gate/delay unit while the unipolar output is connected to an analog/digital board. The calibration was achieved with the 662 keV peak of a $Cs^{137}$ radioactive source. An attenuator connected between the test pulse and the pre-amplifier was used to obtain the pedestal and hence check that the pedestal corresponds to an energy of zero keV in the final A/D board. Analysis of the calibration data gives an energy resolution of approximately 4%.

### 11.3.2 Measurements Results

Results of the measurements at both locations with two beam energies are plotted in Fig. 28. SR and residual gas bremsstrahlung, described in the simulation section, are the two dominant background sources. At $4.5$ GeV in both locations the energy of the photons covers the whole measurement spectrum, only photons from gas residual bremsstrahlung reach the detector. In LOC2 more photons are detected. At 7 GeV in both locations peaks are observed at $n \cdot 60$ keV for integer $n$, which is the signature of SR. However, in LOC1 more photons from this process are observed. A high peak is also found at zero keV in the various figures which correspond to the case where no photons are detected and the acquisition is triggered on by the beam crossing. This is in accordance with Poisson statistics.

### 11.3.3 Comparison Measurements with Simulations

A study of the expected number of events $\mu$ detected for the overall background simulation is performed. This lead to a study of $\mu$ as a function of the effective aluminium beampipe

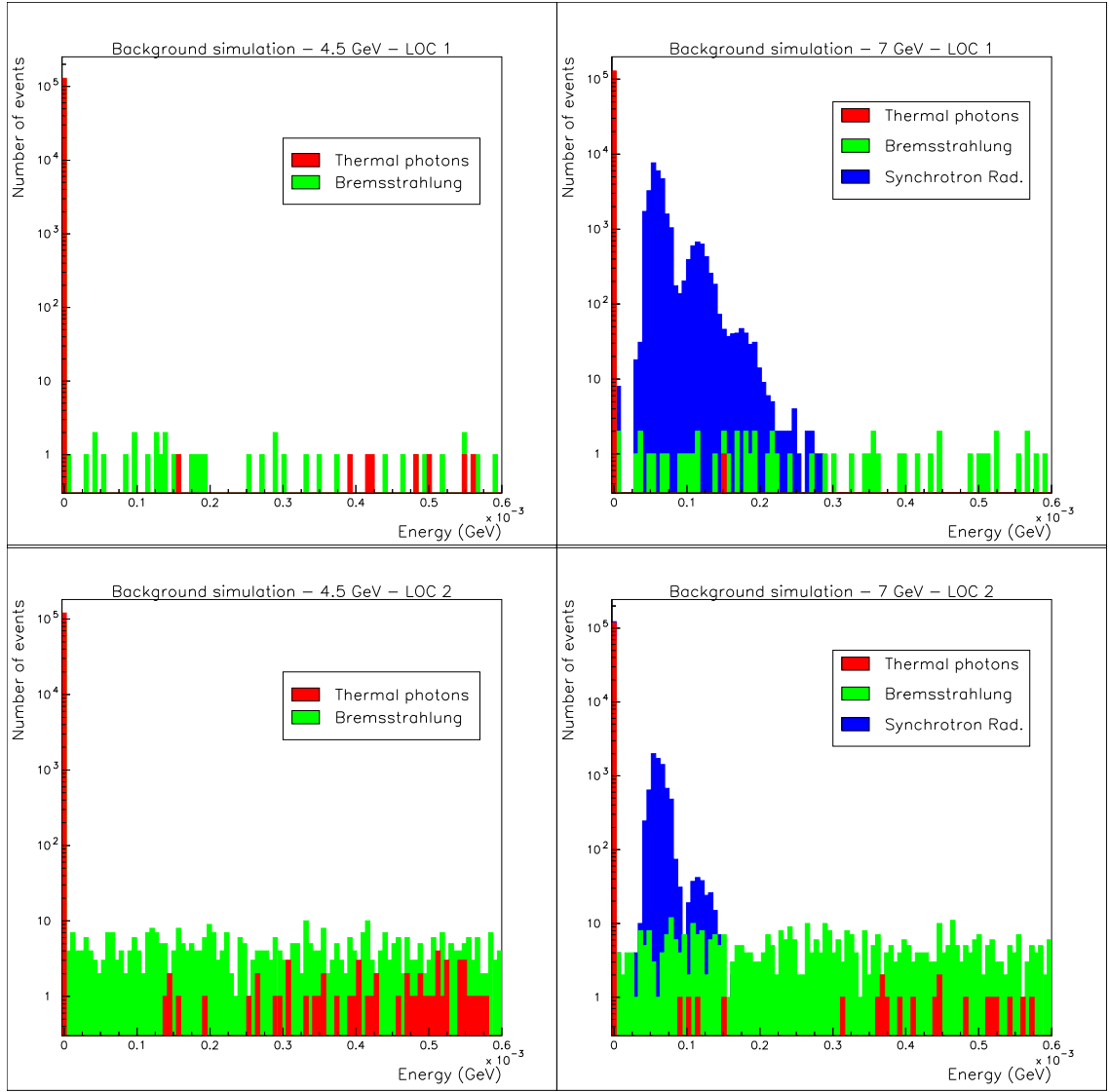

Figure 25: Simulation results for the three background processes under investigations for two beam energies and locations over 1 second observation time (130000 bunches).

thickness which is seen by the photons. For the overall backgrounds and measurements, normalised to the number of bunches $N_{bunch} = 130000$, the expected number of events $\mu$ is calculated using:

$$\mu = -\ln \frac{N_{\gamma_0}}{N_{bunch}} \tag{38}$$

with $N_{\gamma_0}$ the number of events in the zero photon peak. The analysis shows that the photons in location 1 have to go through approximately 145 mm of beampipe material (aluminium) whereas in location 2 the thickness of material is equivalent to 165 mm. For the effective beampipe thickness defined above at location 1 and 2 the number of events can then be computed and compared to the measurements performed over 130000 bunches as shown in Tab.6. A good agreement is observed between the simulation and the measurements for an energy range between 20 and 600 keV. However, it has to be noted that at a beam energy of 4.5 GeV the expected number of events calculated using the peak at zero photons for simulated data is much lower than $\mu$ calculated from the measurements. This indicates that less events have to be expected for measurements over the energy range extended to the maximum event energy in comparison to the simulation calculations and provides an upper

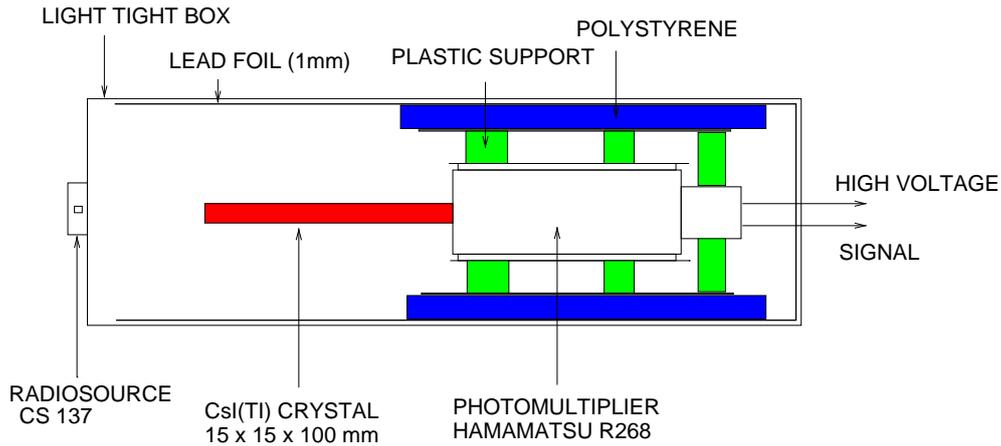

Figure 26: CsI crystal and photomultiplier mounted in a light-tight box.

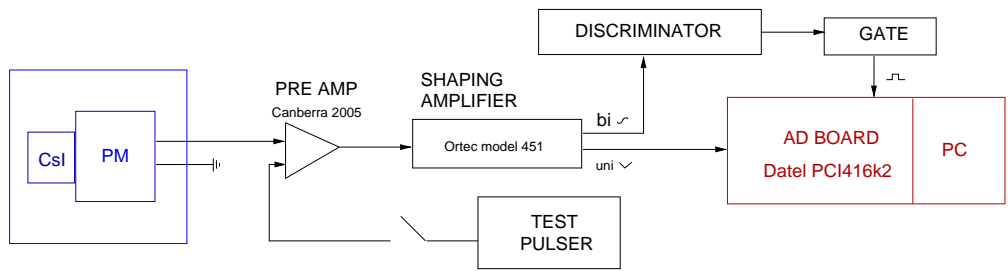

Figure 27: Readout electronics for the CsI crystal.

limit given by $\mu$ for the simulation.

## 11.4   Full Background Simulations

In the next step, the results from the analysis for energies up to 600 keV (Upper energy threshold for the first detector setup) have to be redone for energies up to the actual beam energy in order to accommodate the possibility of an energy detector acceptance with covers the whole energy range. An upper limit would be the Compton peak of the signal at 350 MeV. A material which would enable the measurement of such photons would be PbWO (Lead-tungsten). It is planned to perform a second set of background measurements at the end of 2001 or beginning of 2002. Preliminary results for the full background simulations are shown in Fig. 29.

## 11.5   Signal plus Background

When scanning the electron beam with the laser, several scattered photons are produced. The number of which, $N_{gen}$, can range from a few to thousands. For a variable number of photons produced, simulation of the Compton signal at a beam energy of $4.5$ GeV is studied using an effective aluminium material thickness of $145$ mm. The average energy of the detected photons is $158.2$ MeV and approximately 36% of the photons produced are contained in a cone of $2 \cdot 10^{-3}$ steradians, while the remaining 64% are scattered away due to interactions with the beam pipe material. The collected 35% of the photons contain roughly 80% or the produced energy. The contribution of background is calculated using the algorithm described in the previous section. An example of the resulting spectrum of the energy in the calorimeter including all backgrounds is displayed in Fig.30.

In the following the photon energy entering the detector over each bunch crossing is studied as a function of the number of scattered photons which are generated in the laser beam

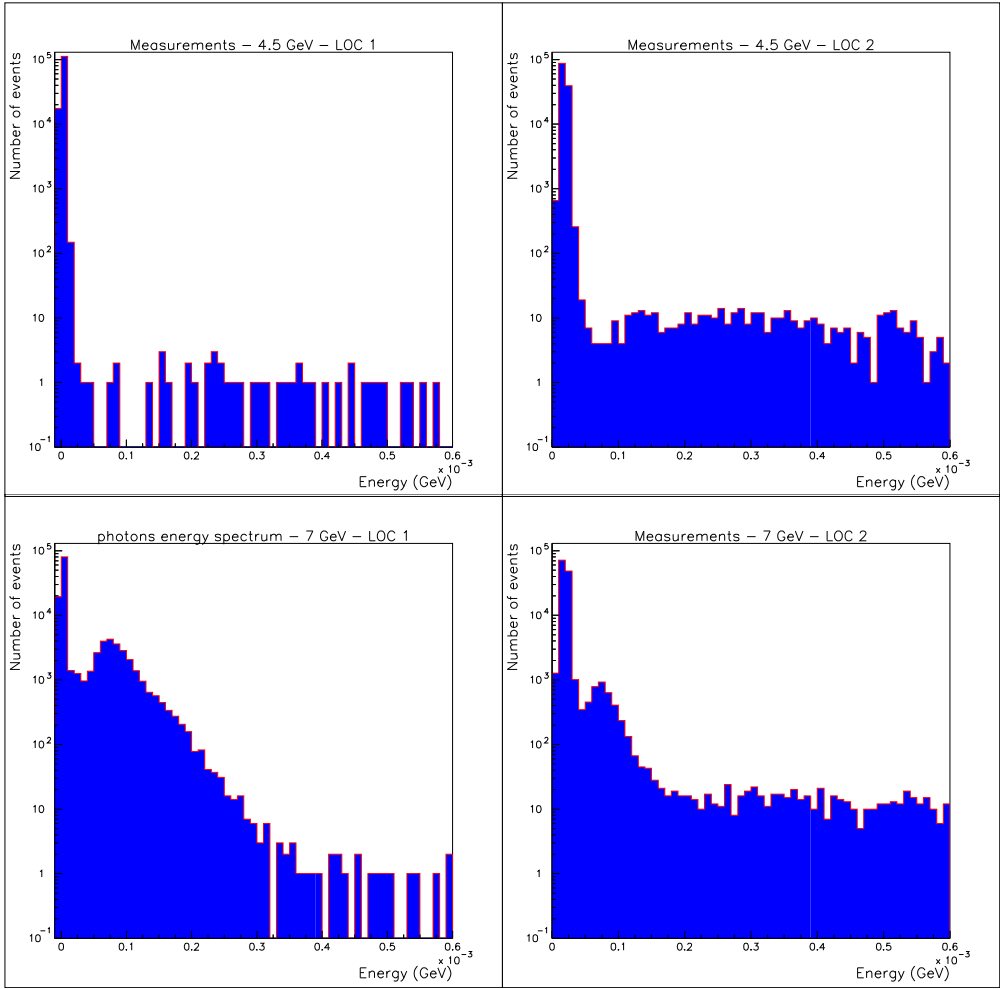

Figure 28: Background measurements data taken over 1 second measurement time corresponding to 130000 bunches.

interaction. This leads to a study of the resolution $\Delta E_{sd}/E_{sd}$ of the smeared signal plus backgrounds distribution, which is a combination of two types of errors: the error on the intrinsic energy spread $E$ determined by the shape of the spectrum in Fig. 30 convoluted with the number of photons produced and the error from a finite resolution of a detector $\Delta E_{cal}/E_{cal}$ as calculated with

$$\frac{\Delta E_{sd}}{E_{sd}} = \sqrt{\left(\frac{\Delta E}{E}\right)^2 + \left(\frac{\Delta E_{cal}}{E_{cal}}\right)^2} \tag{39}$$

or

$$\frac{\Delta E_{sd}}{E_{sd}} = \sqrt{\left(\frac{\Delta E}{E}\right)^2 + \left(\frac{f_s}{E}\right)^2} \tag{40}$$

with $f_s$ the smearing factor of the calorimeter. The overall resolution is plotted in Fig.31. The graph shows that a limit on the energy resolution is obtained for a large number of photons created per interaction: The intrinsic resolution limit is 5.1%. However for low photons number, the accuracy decreases which imposes a lower practical limit on the laser power.

| Location | Energy | Simulation | | Measurements | |
|---|---|---|---|---|---|
| | [GeV] | $\mu$ | $N_{ev}$ | $\mu$ | $N_{ev}$ |
| 1 | 4.5 | $2.29 \cdot 10^{-2}$ | $50 \pm 7$ | $1.57 \cdot 10^{-3}$ | 47 |
| 1 | 7.0 | $2.59 \cdot 10^{-1}$ | $28951 \pm 171$ | $2.47 \cdot 10^{-1}$ | 28321 |
| 2 | 4.5 | $4.22 \cdot 10^{-1}$ | $489 \pm 22$ | $1.70 \cdot 10^{-2}$ | 476 |
| 2 | 7.0 | $6.38 \cdot 10^{-2}$ | $4795 \pm 71$ | $6.37 \cdot 10^{-2}$ | 4790 |

Table 6: Background simulation results of the expected number of events and the number of events $N_{ev}$ over the energy range between 20 keV and 600 keV for 130000 bunches.

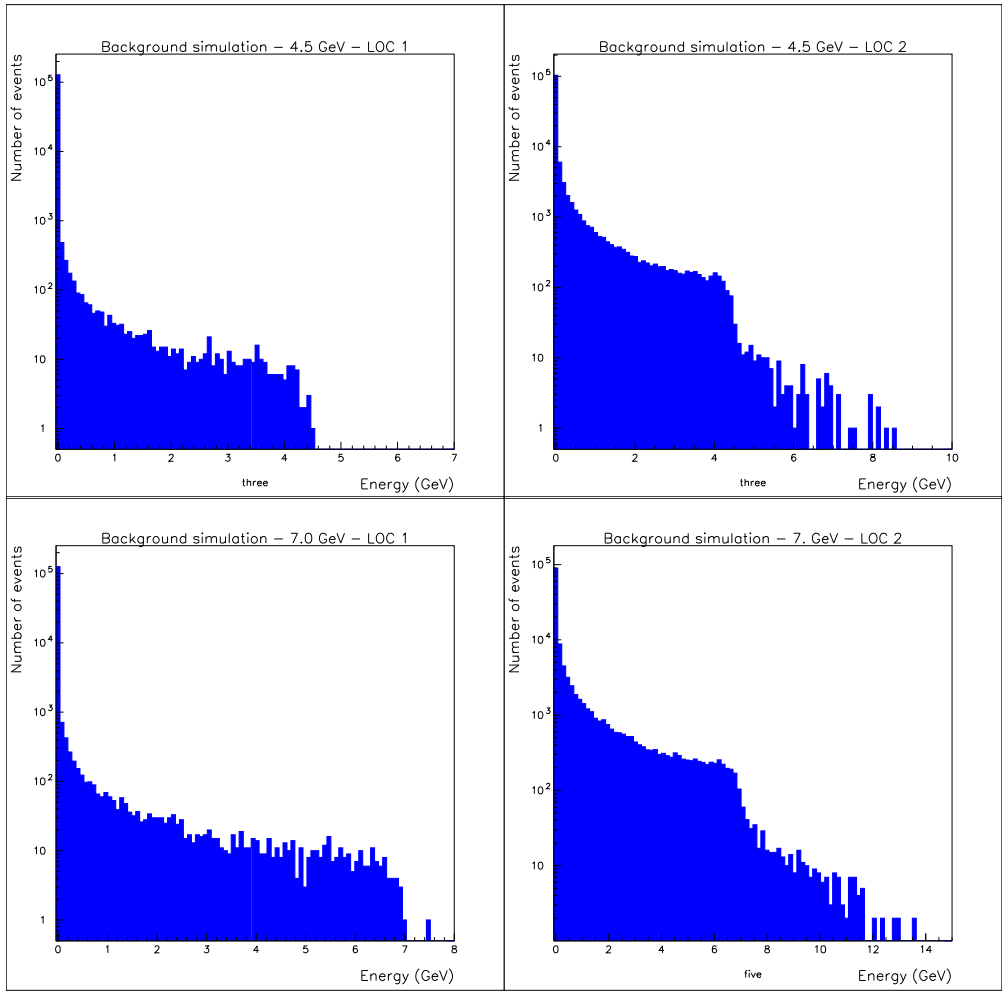

Figure 29: Full background simulations for beam energies up beam energy over 1 second measurement time corresponding to 130000 bunches. No energy resolution from the detector is applied.

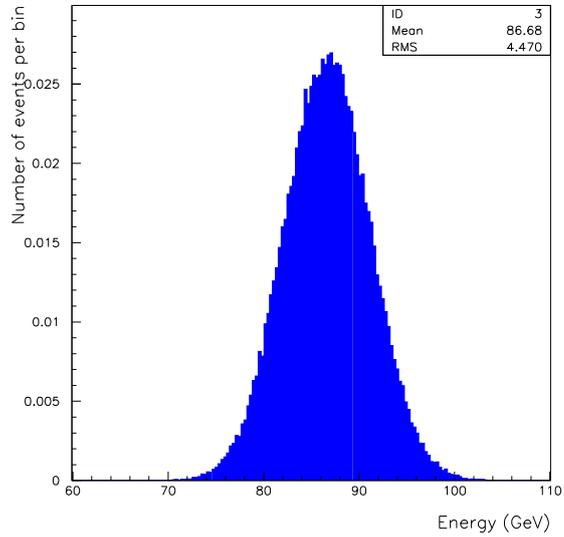

Figure 30: Normalised energy distribution of signal plus backgrounds with an effective aluminium thickness of 145 mm. The error on the energy smearing is 5.1% for $N_{gen} = 1520$.

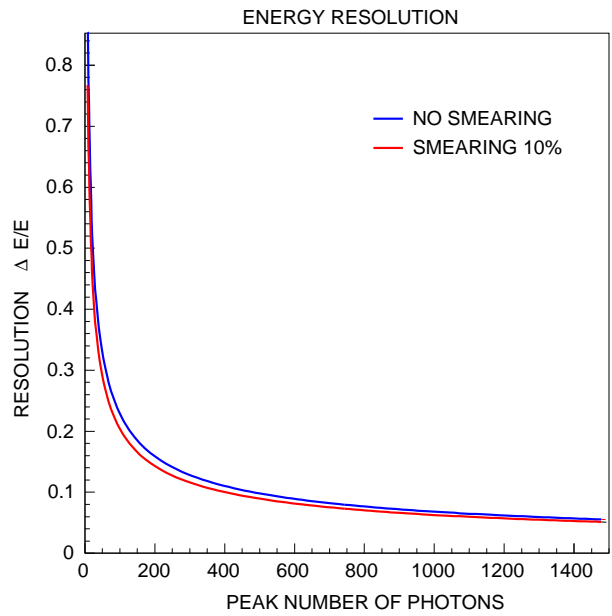

Figure 31: Limits for the energy resolution for photon yields up to 1500 photons.

# PROPOSING A LASER BASED BEAM SIZE MONITOR
# FOR THE FUTURE LINEAR COLLIDER

G.A. Blair[a], J. Frisch[b], K. Honkavaara[c], T. Kamps[a*], F. Poirier[a] I. N. Ross[d], M. Ross[b]
H. Schlarb[c], P. Schmüser[c], S. Schreiber[c], D. Sertore [e], N. Walker[c], M. Wendt[c], K. Wittenburg[c]

[a] Royal Holloway University of London, Egham, Surrey, TW20 0EX, UK

[b] Stanford Linear Accelerator Center, Stanford, CA 94309, USA

[c] Deutsches Elektron–Synchrotron DESY, D-22603 Hamburg, Germany

[d] Rutherford Appleton Laboratory, Chilton, Didcot, Oxon OX11 0QX, UK

[e] INFN Milano LASA, I-20090 Segrate (MI), Italy

## Abstract

Compton scattering techniques for the measurement of the transverse beam size of particle beams at future linear colliders (FLC) are proposed. At several locations of the beam delivery system (BDS) of the FLC, beam spot sizes ranging from several hundreds to a few micrometers have to be measured. This is necessary to verify beam optics, to obtain the transverse beam emittance, and to achieve the highest possible luminosity. The large demagnification of the beam in the BDS and the high beam power puts extreme conditions on any measuring device. With conventional techniques at their operational limit in FLC scenarios, new methods for the detection of the transverse beam size have to be developed. For this laser based techniques are proposed capable of measuring high power beams with sizes in the micrometer range. In this paper general aspects and critical issues of a generic device are outlined and specific solutions proposed. Plans to install a laser wire experiment at an accelerator test facility are presented.

## 1 MOTIVATION

High luminosity is the key to many of the physics processes of special interest at the Future Linear Collider [1]. This fundamental point is the main physics motivation for this project and justifies considerable efforts to ensure that the accelerator can deliver on its excellent luminosity potential. The case for the highest luminosities is now globally accepted and all the Linear Collider proposals currently have this as their goal, with quoted luminosities of a few $\times 10^{34}$ cm$^2$s$^{-1}$. The key motivation for this project is to add to the arsenal of tools that the machine will need to maximize its luminosity performance. In particular this project aims to provide a reliable and flexible method of obtaining real-time information on the emittance and quality of the beam and hence to allow feedback for maximizing the luminosity.

## 2 EMITTANCE MEASUREMENT

In this project we limit our attention to the measurement of the electron beam transverse phase space (transverse emittance) because it is the fundamental determining factor for the final transverse beam-spot size at the interaction point (IP). It is important to keep the emittance low so as to maximize the luminosity at the IP and much effort is spent in designing the accelerator and beam delivery system (BDS) to avoid sources of emittance growth. The BDS generically consists of approximately a kilometer of beam optics providing collimation, chromatic correction and final focusing. There are many potential sources of emittance growth which in general will be time dependent and will require continuous measurement and feedback to correct.

The aim is to measure the emittance of the beam to better than 10% as it approaches the IP and this will require a number of profile measurements with the same precision along the BDS. In Tab. 1 beam profile parameters for FLC designs are listed.

| | | CLIC | NLC/JLC | TESLA |
|---|---|---|---|---|
| BDS | $\sigma_x[\mu m]$ | 3.4 to 15 | 7 to 50 | 20 to 150 |
| | $\sigma_y[\mu m]$ | 0.35 to 2.6 | 1 to 5 | 1 to 25 |
| IP | $\sigma_x^*[nm]$ | 196 | 335 | 535 |
| | $\sigma_y^*[nm]$ | 4.5 | 4.5 | 5 |

Table 1: Beam spot sizes for various Linear Collider designs. Quoted are numbers for CLIC [2], NLC/JLC [3], and TESLA [4].

A set of transverse profile measurements at several points along the beam line separated by a sufficient betatron phase advance can be translated into a determination of the emittance. At least four scanning stations will be required for each lepton beam, possibly fired by a single laser system plus laser beam transportation. Each station will need to provide a profile along three directions, as required to specify an ellipse. Relating a set of such transverse profiles to the emittance and optimizing the layout of scanning stations within a BDS design will form an interesting parallel

project, that will be addressed via detailed simulations.

The electron bunch transverse profile has been measured in the past by intersecting the electron beam with a solid wire and by counting the subsequent background rate as a function of the relative position of wire and bunch. Using this technique, resolutions of typically a few $\mu$m can be obtained, at the expense of some disruption to the beam. This technique cannot be used universally at the LC, however, because the beam-spot sizes can be much smaller, the need for continuous measurement precludes an invasive technique and the intensities are so great that the wires would be quickly damaged, even if swept rapidly through the beam. For these reasons, it is necessary to develop a novel technique that can run continuously and reliably during machine operation, that does not get destroyed by the beam and that can be sufficiently fast so as to be sensitive to individual electron bunches within the bunch train. All these advantages could in principle be provided using optical scattering structures.

## 3  COMPTON SCATTERING

The basic idea is to replace a solid wire in a beam profile scan by a narrow beam of laser light. The fundamental process at work is then the Compton effect, where photons are scattered out of the laser beam by the incoming electrons. By counting the rate of scattered photons (or electrons) as a function of the relative position of the laser waist and the electron beam, a measurement of the bunch transverse profile can be obtained. The relativistic Compton process has been studied in detail elsewhere [5]. Relevant aspects for our analysis together with detailed simulations are collected in [6].

Several schemes have been proposed to use optical scattering structures to serve as diagnostics to measure the bunch length and the beam profile [7].

### 3.1  OPTICAL SCATTERING STRUCTURES

Common to all optical scattering structures is that they must have features smaller or similar in size to the particle beam under measurement. Several types of laser spot structures can be generated with common optical setups. In the following some optical structures are listed together with their performance rating:

**Laser wire (gaussian profile)** The laser beam is here focused to a small gaussian spot with radius $\omega_o$. If we consider a diffraction limited, finely focused beam waist, the minimal achievable spot radius is given by $\omega_o = \lambda/(\pi\theta)$, where $\lambda$ denotes the laser wavelength and $\theta$ the half opening angle of the laserbeam at the waist (see Fig.1). The distance over which the laser beam diverges to $\sqrt{2}$ of its minimum size is called the Rayleigh range $x_R$ and defines the usable length of the laser wire. The smallest achievable spot size with diffraction limited optics is in the order of $\omega_o \sim \lambda$. With Nd:YLF or YAG laser working at higher harmonics electron spot sizes from $\sigma_y > 350$nm

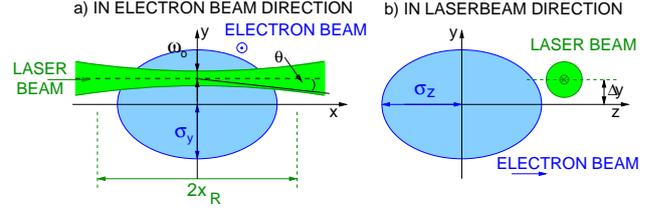

Figure 1: Scheme of a gaussian laser beam focused to its diffraction limit scanned over an electron beam.

can be measured with high accuracy. The laser beam power must be in the order of a couple of MW to yield a few thousand Compton photons per scan spot. Critical issues of a laser wire design are the diffraction limited optics, which must withstand such a high beam power and the scanning system, enabling intra-train scanning of consecutive bunches.

**Laser wire (dipole mode)** The resolution of the laser wire can be enhanced by generating an artificial transverse dipole mode by means of a lambda half waveplate, where half of the gaussian is shifted in phase by $90^o$. Such a waveplate can easily be installed in the optical path of the laser wire and would enhance the resolution of the device by roughly a factor of two aiming at beam sizes in a region from 250nm $< \sigma_y <$ 500nm.

**Laser Interferometer** Towards beam sizes in the nanometer range, a standing wave interference pattern generated by crossing two laser beams has been proposed and successfully tested at the FFTB experiment [8]. The fringe spac-

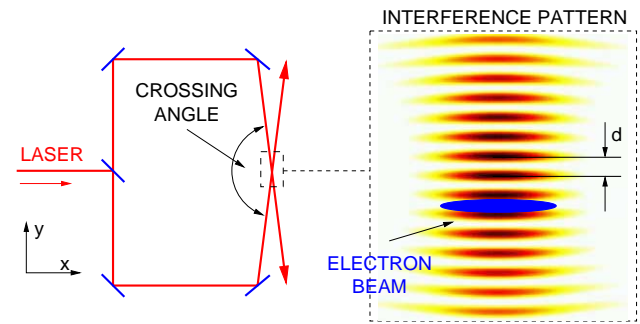

Figure 2: Scheme for the generation of an interference pattern with fringe spacing $d$.

ing of the interference pattern (see Fig. 2) depends on the laser wavelength and on the crossing angle. The electron beam is moved over the pattern and the Compton scattered photons are detected. If the beam size is small compared to the fringe spacing, a modulation of the Compton signal is observed which is proportional to the transverse electron beam size. This modulation vanishes if the beam size is large compared to the fringe spacing. The smallest observed spot size with this technique was about 58 nm [9].

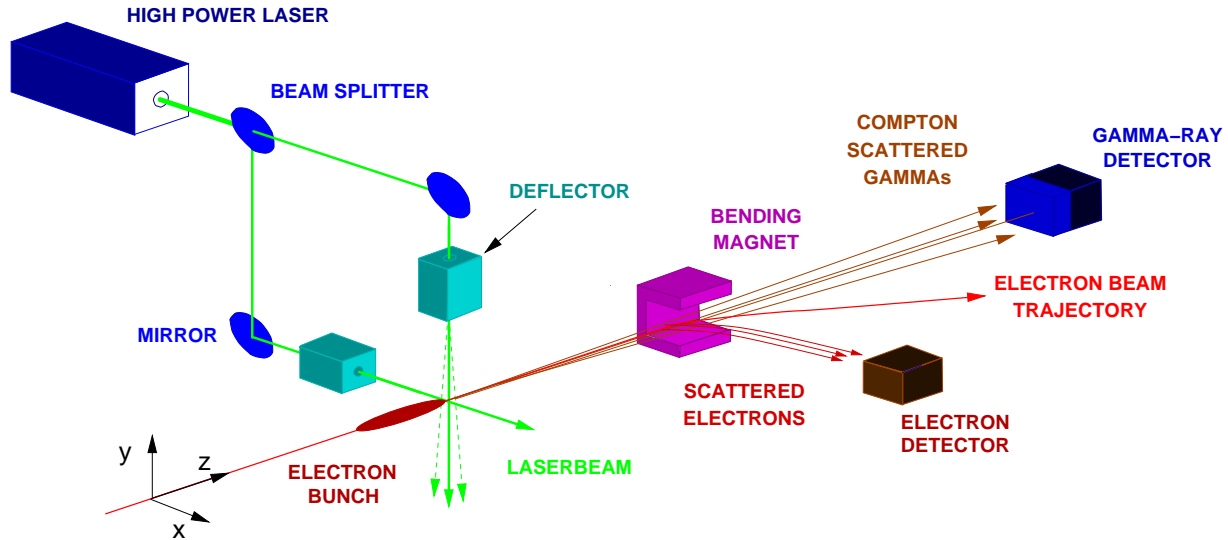

Figure 3: Schematic setup for a laser wire beam profile monitor.

## 4   LASER WIRE SETUP

The setup for a laser wire beam profile monitor is sketched in Fig. 3. A high power laser beam is divided into two different optical paths for scanning the horizontal and vertical beam size. The scanning is foreseen to be done either with piezo-driven mirrors or with acousto-optic scanners. Before the interaction with the electron beam the laser beams are focused. The electron beam is then bent away while the Compton scattered photons travel along a straight line where they are detected with a calorimeter. Scattered electrons will be bent more strongly than particles with the nominal beam energy enabling detection at a location after the bending magnet.

## 5   TEST OPTIONS

It is planned to install a complete laser wire scanner at the PETRA accelerator at DESY in summer 2002. While beam sizes at PETRA ($10 - 100\ \mu m$) are comparable with typical FLC BDS numbers, the energy of the electron beam is lower in the range from $4.5 - 12$ GeV. Recent results from background measurements show [6] that a sufficient signal to noise ratio can be reached even with a medium power laser with peak power less than $10$ MW. In Spring 2002 tests of subsystems of a laser wire are planned at CTF2/3.

## 6   CONCLUSIONS

It is anticipated that laser wire scanners will be the standard beam size instrumentation tool for the beam delivery system of all FLC designs. First design ideas exist with the prototype setup tested at SLC/SLD [10]. Our aim is to elevate this design to a compact, non-invasive device where a high-power pulsed laser is scanned across the electron beam with novel scanning techniques.

# SIMULATION STUDIES AND BACKGROUND MEASUREMENTS FOR A LASER BASED BEAM SIZE MONITOR FOR THE FUTURE LINEAR COLLIDER

K. Balewski[a], G.A. Blair[b], T. Kamps[b*], F. Poirier[b], K. Wittenburg[a]

[a] Deutsches Elektron–Synchrotron DESY, D-22603 Hamburg, Germany

[b] Royal Holloway University of London, Egham, Surrey, TW20 0EX, UK

## Abstract

At several locations of the beam delivery system (BDS) of a future linear collider (FLC), beam spot sizes ranging from several hundreds to a few micrometers have to be measured. It is anticipated that laser wires will be used for this task in any FLC design. In order to optimize a laser wire system, simulations and background measurements have been carried out. Results are presented from simulations of the Compton scattering for the PETRA scenario. Furthermore results from measurements of backgrounds like synchrotron radiation and gas scattering have been measured at the positron storage ring PETRA at DESY and will be discussed.

## 1 INTRODUCTION

Laser wire scanners (LWS) will play an increasing role as the standard tool for beam size measurements [1] in the micrometer range for the BDS at any FLC design [2]. The LWS operation principle is based on the interaction between the electron beam and a finely focused laser beam. The electron bunch transverse size is measured by scanning the electron beam with the laser beam. Photons are then scattered out of the laser beam by Compton scattering. By counting the rate of Compton photons (or degraded electrons) as a function of the relative position of the laser waist and the electron beam, a measurement of the transverse beam size can be obtained. To specify the required laser beam power and detector performance, simulations and measurements of the backgrounds for such a measurement are of fundamental importance. In the following calculations of the signal level as well as of possible background sources are given. Furthermore results from background measurements in the relevant accelerator environment are discussed.

## 2 COMPTON SCATTERING

The fundamental process at work at the LWS is Compton scattering. In the classical limit with photon energies smaller than the electron energy and with electrons at rest, photon electron scattering is described by Thomson scattering. For high-energy photons non-classical effects must be taken into account leading to Compton scattering. With moving electrons the process is called inverse Compton scattering, where the moving electrons transfer energy to the photons yielding substantial fluxes of photons in the optical to X-ray region [3]. For the LWS, the total energy of the scattered photons per electron bunch laser pulse crossing is considered as the signal process. The number of photons $N_C$ is directly proportional to the laser beam power $P_L$ and wavelength $\lambda$ according to [4]

$$N_C = N_b \frac{P_L \sigma_C \lambda}{c^2 h} \frac{1}{\sqrt{2\pi}\sigma_s} \exp\left(\frac{-y^2}{2\sigma_s^2}\right) \qquad (1)$$

where $N_b$ is the number of electrons in a bunch, $y$ the relative offset between laser and electron beam and $\sigma_s^2 \equiv \sigma_y^2 + \omega_o^2$ the overlap region. The electron beam size is $\sigma_y$ and the laser beam waist at the interaction point is $\omega_o$. The Compton cross section $\sigma_C$ is in Fig. 1 evaluated for two scenarios: One for a typical linear collider test facility

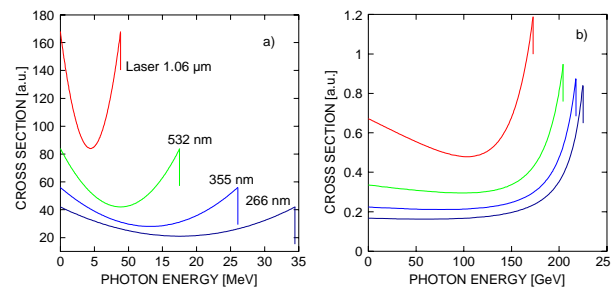

Figure 1: Compton cross section for the first four harmonics of an Nd:YAG laser scanning a 1 GeV (a)and 250 GeV (b) electron beam.

beam energy (1 GeV), where sub-systems of a LWS will be tested and for a typical linear collider beam delivery system energy (250 GeV). Currently a Nd:YAG based laser system is the instrument of choice because of its performance capabilities with respect to high power and small spot size.

## 3 BACKGROUND SOURCES

For the use of a LWS, the background conditions at the detector locations are important. The biggest source of background photons in the low energy region is synchrotron radiation emitted by the electron beam in bending magnets. The spectrum of synchrotron radiation is characterized by the critical energy which is in the keV region for beam energies of several GeV. Another background source arises from beam gas scattering. Here the electrons

---

in the beam interact with the residual gas. Bremsstrahlung is the dominant process at high energies with a cross section combined from the individual cross sections from photon emission at the nucleus and emission at the bound electron. The bremsstrahlung spectrum for emitted photons with energy $E_\gamma$ goes up to the beam energy $E_b$ proportional to $(E_\gamma/E_b)^{-1}$. Both spectra, for synchrotron radiation and bremsstrahlung are plotted in Fig. 2.

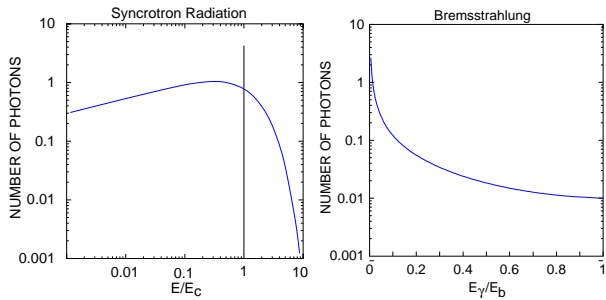

Figure 2: Photon spectrum from synchrotron radiation and bremsstrahlung.

## 4 SIMULATIONS

The aim of the simulation studies is a full Monte Carlo simulation of the signal (Compton) and all relevant background processes. This includes the modeling of the complete measurement setup, from the scattering process to the analog-digital-converter (ADC) of the readout electronics of the detector. This is of fundamental importance for the specification of the laser system and the detector. So far the Compton process is modeled in a realistic accelerator environment including beampipe, magnets, and vacuum windows. The parameter set for the electron beam are for the PETRA positron storage ring at DESY in Hamburg, because prototype tests of a LWS are foreseen with this machine. The simulation work is carried out in the Geant4 [5] framework. The standard toolkit is used for multiple scattering while the low energy electromagnetic toolkit [6] is includes Compton and Rayleigh scattering, photoelectric effect, bremsstrahlung, and ionization at energies 250 eV. A specific Monte Carlo generator [7] is implemented for synchrotron radiation photons. In Fig. 3 simulation results are shown for the Compton process with PETRA accelerator parameters. The low energy peak in the spectrum is caused by multiple scattering processes at low energy in the 2 mm steel vacuum window. Full Monte Carlo simulations of the background processes are under way.

## 5 BACKGROUND MEASUREMENTS

Preliminary background measurements have been performed at the positron storage ring PETRA. Two locations were used for the measurements at two different energies. In Fig. 4 the measurement setup is depicted. A Cs(I) crystal ($1.5 \times 1.5 \times 10$ cm$^3$) mounted to a photomultiplier was

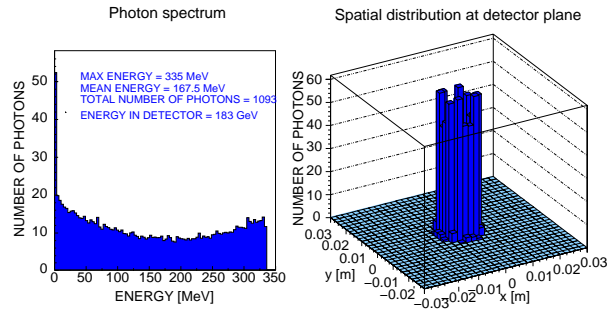

Figure 3: Photon spectrum of Compton photons simulated with Geant4. Scattering of a 10 MW peak power green ($\lambda = 532$ nm) laser beam with 1 $\mu$m spot size at a 4.5 GeV electron beam with transverse beam size of 100 $\mu$m. For the spatial distribution calculation the detector positioned 4 m away from the interaction.

used. The crystal and the photomultiplier are packed together with a support structure in a light tight box made of lead. An aperture is drilled on the side facing the electron beam. The whole package is taped with black tape and positioned at the two locations tangent to the beampipe and 30 cm away from the dipole magnets to avoid any electromagnetic effects on the detector. Before making the measurements the detector together with its readout electronics were calibrated in terms of energy using a radioactive source. The beam parameters of the positron beam at PETRA relevant for the measurements are gathered in Tab. 1 Results from measurements at the two locations with two

| Beam energy | 4.5 and 7 | GeV |
|---|---|---|
| Beam current | 1.55 to 1.77 | mA |
| Particles per bunch | 7.5 to 8.5 | $10^{10}$ |
| Repetition rate | 130 | kHz |
| Bending angle | 28.08 | mrad |
| Vacuum pressure | 1 and 2 | $10^{-10}$mbar |

Table 1: PETRA beam parameters relevant for background measurements.

beam energies are plotted in Fig. 5. Synchrotron radiation and bremsstrahlung are the two dominant background sources. The energy of the photons from synchrotron radiation at 4.5 GeV is too low ($E_C \simeq 1$ keV) to pass the beampipe material. At 7 GeV the background is a superposition of synchrotron radiation with strong components of bremsstrahlung. At location two, after the long straight section, the electrons have to pass through about 16 times more residual gas, which increases the rate of bremsstrahlung by that amount.

## 6 CONCLUSIONS

Laser wire scanners (LWS) will play an increasing role as the standard tool for beam size measurements at future linear collider. To fully exploit the potential of such a de-

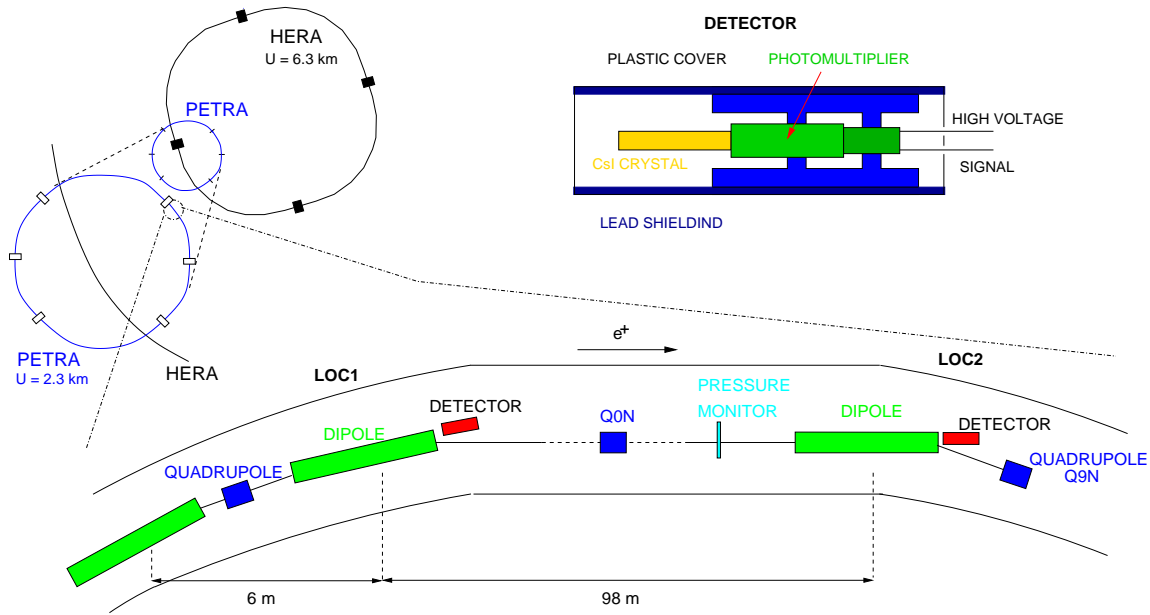

Figure 4: Schematic setup for background measurements at PETRA.

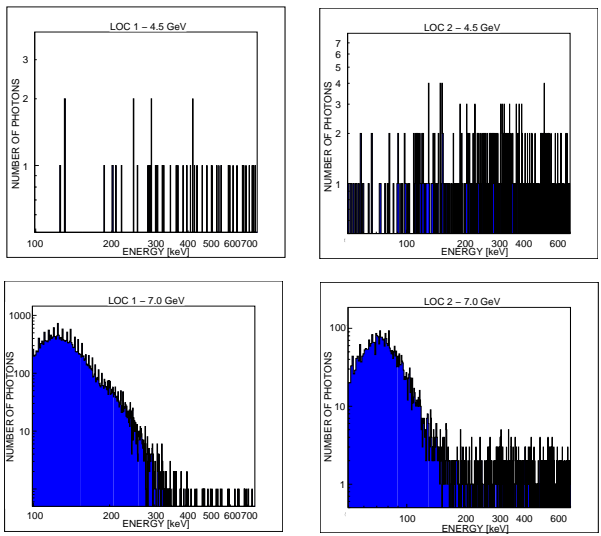

Figure 5: Background spectra measured at PETRA. Data was taken over a 1 sec long period for each measurement.

vice, studies are under way to supply full Monte Carlo simulations of all relevant processes connected with the LWS. Furthermore background measurements at the PETRA accelerator have been performed, enabling laser and detector specification for a system test of a LWS. A much faster detector, made of lead tungstate, is under study. It might enable a 10 MHz sampling rate which is the nominal bunch rate at PETRA.

# R&D TOWARDS A LASER BASED BEAM SIZE MONITOR FOR THE FUTURE LINEAR COLLIDER

G.A. Blair[a], T. Kamps[a*], H. Lewin[c], F. Poirier[a], S. Schreiber[c], N. Walker[c], K. Wittenburg[c]

[a] Royal Holloway University of London, Egham, Surrey, TW20 0EX, UK

[c] Deutsches Elektron–Synchrotron DESY, D-22603 Hamburg, Germany

## Abstract

Laser wires will play an important role as the standard monitor for beam size measurements with micrometre resolution for the beam delivery system at any future linear collider. Some R&D work is still necessary to elevate preliminary laser wire designs to a compact, non-invasive and fast-scanning device. In this paper the latest R&D together with recent measurements and simulations are presented. Schemes to measure the beam size in a bunch train and from train-to-train are presented together with an evaluation of scanning techniques meeting these requirements. Results from simulations and measurements with a laser focus system and of a proposed Compton calorimeter are reported. Furthermore plans are outlined for the installation of a fast laser wire experiment at the PETRA accelerator at DESY.

## 1 INTRODUCTION

The principle of laser wire operation is illustrated in Fig. 1, where light from a laser is focused down to a small spot and scanned across the incoming electron beam. The resulting Compton-scattered photons are detected downstream and the measurement of the total energy of these photons as a function of the laser spot position yields the electron bunch transverse dimensions. In the following, the various components are discussed in more detail and progress towards building a working system is reported.

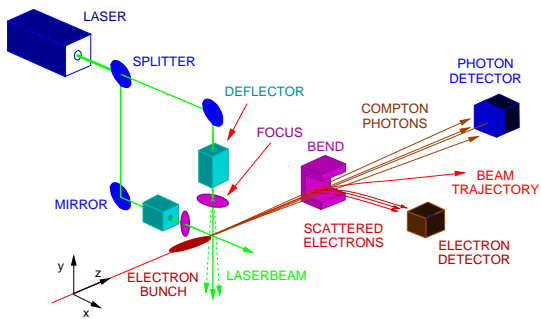

Figure 1: A generic laser wire profile monitor.

## 2 FOCUSING OPTICS

For a gaussian laser beam the RMS spot size at the interaction should be smaller than the electron beam size

* t.kamps@rhul.ac.uk

$\sigma_o \leq \sigma_y$. At PETRA the electron beam has dimensions of $\sigma_y = 20 - 30$ $\mu$m and $\sigma_x = 200 - 300$ $\mu$m leading to $\sigma_o = 5 - 10$ $\mu$m and a Rayleigh range of at least $z_R \geq 300$ $\mu$m in order to accommodate completely the horizontal beam size. High laser peak power between $P_{max} = 1 - 10$ MW is necessary to obtain a good signal-to-noise ratio [1]. Thus the laser optics has to withstand this amount of power and, for flexibility, should be able to work with at least the first two laser harmonics. The back

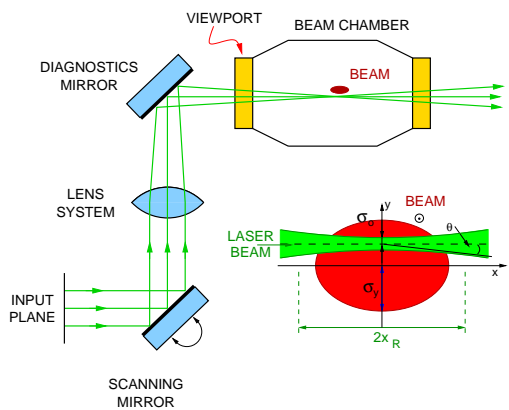

Figure 2: Layout of scanning and focusing system.

focal length should be $BFL \simeq 150$ mm to incorporate scanning and diagnostic mirrors (see Fig. 2). In order to conform with the above requirements, an air-spaced achromatic laser objective with three lenses was chosen.

The laser beamline around the focusing triplet was simulated using the ZEMAX code [2]. To first order (neglecting abberations) the minimum spot radius is determined by the f-number $f\#$ of the lens, the mode quality $M^2$ and the laser wavelength $\lambda$ according to $\omega_o \approx M^2 \lambda f\#$. The simulation code also allows for higher orders from spherical abberations and coma.

### 2.1 Measurements

Spot size measurements with the proposed lens triplet were performed in order to study the focusing, beam propagation around the best focus, and tolerances. Knife edge scanning with a piezo driven razor blade was chosen as the measurement technique because of its simplicity and high precision. This technique involves passing a blade between the laser beam and a photo detector and measuring the light intensity as a function of blade position, thereby providing the integrated laser beam profile.

In Fig. 3 the measurement setup is sketched. The laser light coming from a green HeNe laser is guided with a mirror into a Keplerian telescope for collimation. At the

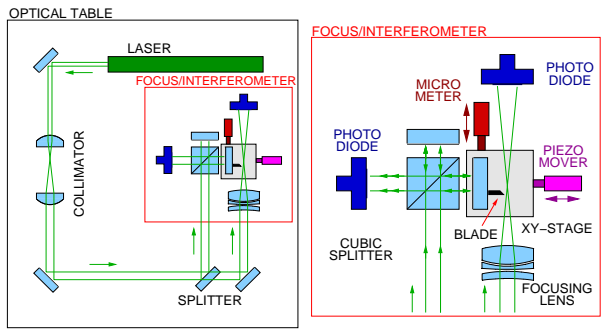

Figure 3: Laboratory setup to measure small laser spot sizes using the knife edge technique.

beam splitter half the beam power is guided into a Michelson interferometer while the other half passes another mirror before going through the focusing lens. Following this lens the focused beam is collected by a photo diode. The blade together with its interferometer mirror are moved by a piezo driven actuator. The step width of the piezo actuator is monitored constantly in the interferometer arm. For one complete scan of the beam profile and propagation several slices of the laser beam were measured in forward and reverse direction (see Fig. 4a for an example of one slice). This was repeated with a viewport window in the beam path between the focusing lens and the knife edge. The measured propagation of the beam envelope is shown

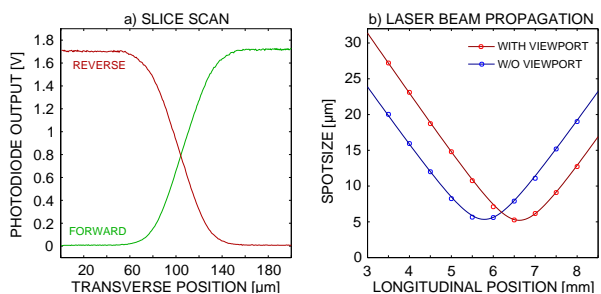

Figure 4: Left: One slice (measured forward and reverse) of the laser beam around the best focus. Right: RMS spot propagation around the best focus.

in Fig. 4b. The minimum RMS spot size was obtained by a least square fit using the beam propagation model for gaussian beams [4]. The measured minimum RMS spot size is $\sigma_{o,w} = (5.34 \pm 0.02_{stat} \pm 0.07_{sys})$ $\mu$m with and $\sigma_{o,w/o} = (5.21 \pm 0.03 \pm 0.07)$ $\mu$m without viewport window. The input beam radius is $\sigma_{in} = (1.23 \pm 0.02)$ mm. All measurements agree with numerical simulations and, as expected from these simulations, the spot size is essentially unaffected by the presence of the viewports. The main effect of the viewport is to shift the waist by $\Delta z = (0.83 \pm 0.01 \pm 0.08)$ mm, which is a third of the viewport thickness of $t = 2.50$ mm. The main contributions to

the systematic error are from diffraction effects at the razor blade, which will be tackled in future measurements using a collecting lens.

# 3 SCANNING

## 3.1 Requirements

The total scan range $d$ should be in the order of $d = 10 \times \sigma_y$ the beam size under measurement (for TESLA the vertical beam size is $\sigma_y = 1 - 25$ $\mu$m, for PETRA $\sigma_y = 20 - 30$ $\mu$m). The minimum step width between two scanning points $\Delta d$ and therefore the scan resolution is anticipated to be $\Delta d = \sigma_y/5$ to $\sigma_y/2$. Furthermore the scanner should preserve the mode quality of the laser beam, withstand the high laser peak power and be able to operate over long periods of time. Most importantly, the scanner must match the timing of the macro pulses delivered by the accelerator. TESLA produces bunch trains of 950 $\mu$s length with 2820 bunches each spaced by 337 ns and with a repetition rate of 5 Hz [5]. PETRA as a storage ring can be operated with any harmonic bunch spacing of the repetition rate of 130 kHz. For TESLA it is planned to scan the beam profile with at least ten scan points within one bunch train, which sets the minimum operation frequency of the scanner to 1 kHz and for the laser to 10 kHz. This serves as a guideline for the choice of scanner and for the tests at PETRA.

## 3.2 Candidate Technologies

There are two promising candidates: Acousto-optic (AO) scanners and piezo driven mirrors. AO scanners using Bragg reflection in a block of fused silica are able to operate at very high speed with random access times in the order of $\Delta t = 0.5$ $\mu$s enabling the scanning of every third bunch within TESLA parameters. The devices are also very compact and widely used in industry as Q-switches for high power lasers. The drawback of AO scanners is their low damage threshold and their need for anamorphic beam compression and expansion to match the laser beam profile into the scanner aperture. In addition, the mode quality is dramatically decreased with a diffraction efficiency in the order of 40% for full deflection.

The second interesting technology is based on piezo driven platforms, where a laser mirror is moved by a small stack of piezo electric material sandwiched in a tilting platform. These platforms are able to operate in discrete and continuous mode with frequencies up to 5 kHz within specifications. Since these platforms deflect the laser beam with a mirror, the damage threshold is rather high and the beam distortion should be minimal.

Due to its high damage threshold and the versatility in operation mode, the first tests will be performed with a piezo tilting platform. Preparations are currently under way to perform spot size measurements during high frequency scanning and to quantify any resulting beam distortion.

## 4 DETECTOR

At every laser and electron bunch crossing, a burst of Compton scattered photons is released. The total scattered energy at each burst can then be used to determine the relative position of the laser and electron beams. A series of simulations and measurements are currently under way to determine the most suitable detector for measurements at PETRA and TESLA. The most stringent requirements at PETRA are imposed by the bunch separation (192 to 480 ns) and by space constraints at the beamline.

To avoid pile-up of events, the detector must have a decay time that is short relative to the bunch spacing, so a fast material is required. The material must also have a relatively high scintillation light output, be radiation hard and should have a small radiation length in order to contain fully the electromagnetic shower. The Compton photons are emitted within a small angle relative to the electron direction and so the active volume of the detector must be compact (i.e. possess a small Molière radius) so that it can fit close to the beampipe.

These requirements for the PETRA laser wire calorimeter have led to the choice of lead tungstate (PbWO4) as active material, a crystal whose characteristics [6] are listed in Tab. 1.

| | | |
|---|---|---|
| Radiation length | [mm] | 8.90 |
| Molière Radius | [mm] | 22 |
| Density | [g/cm$^3$] | 8.28 |
| Avg. #Photoelectrons/MeV | | 16 |
| Decay time | [ns] | 5–15 |

Table 1: Relevant PbWO4 characteristics.

A primary requirement of the PbWO4 calorimeter is that it must contain most of the shower resulting from the Compton scattered photons. The overall dimensions required for the crystal in the PETRA case were determined by detailed simulations within the Geant4 [7] framework, using a cuboid shaped crystal of variable length and width. In these simulations photons of energy 350 MeV, which is the maximum energy of a Compton-scattered photon from a 4.5 GeV electron beam, were projected towards the detector. The resulting relative energy containments for various crystal dimensions are shown in Fig. 5a. It can be seen that more than 90% of the incoming photon energy is contained within the crystal for an overall size of 54 mm in width and 150 mm in length, leading to the choice of a 3 by 3 matrix of crystals, each with dimensions $18 \times 18 \times 150$ mm. The energy resolution of such a matrix is presented in Fig. 5b as a function of the number of Compton-scattered photons for three beam energies relevant to the PETRA environment [3]. These simulations show that an energy resolution of better than 5% should be reached with a nominal 1000 Compton-scattered photons at a PETRA beam energy of 12 GeV.

These simulations will be tested against measurements

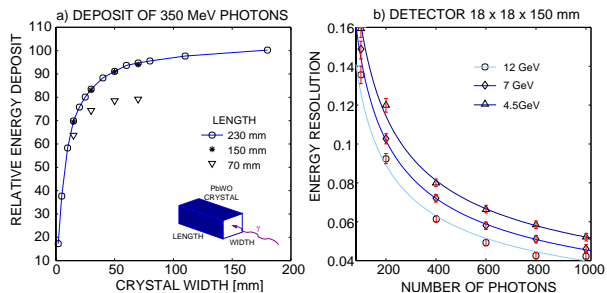

Figure 5: a) Relative energy containment for 350 MeV photons inside a PbWO crystal for three crystal lengths. b) Energy resolution for three beam energies.

using a PbWO4 crystal matrix at the DESY II testbeam, where detailed calibration and efficiency studies are planned.

## 5 SUMMARY

A broad range of studies are underway, aiming towards the installation of a prototype laser wire system at the PETRA storage ring. Detailed design studies, measurements and simulations have been performed for the final focus optics and the implementation of a fast scanning system based on piezo driven mirrors is currently under study. The vacuum chamber for the laser wire is now being constructed at DESY. Detailed simulations of the Compton calorimeter have been performed and test-beam measurements are imminent. Tests of the full scanning laser wire system are expected to take place in 2003.

We acknowledge discussions and help from I. N. Ross, M. Ross and J. Frisch. GAB and TK acknowledge support from the Royal Society and the British Council.

# LASER WIRE SCANNER DEVELOPMENT ON CTF II

J. Bosser, H.H. Braun, E. Bravin, E. D'amico, S. Döbert, S. Hutchins,
T. Lefèvre[*], R. Maccaferri, G. Penn, CERN, Geneva, Switzerland
G.A. Blair and T. Kamps, RHUL, London, England

## Abstract

A laser wire scanner is under development at CERN in the framework of the Compact LInear Collider study (CLIC). A first test has been carried out at the CLIC Test Facility II (CTF II) with the aim of developing a beam profile monitor for a low energy, high charge electron beam. In our set-up a 2.5 mJ, 1047 nm, 4 ps laser pulse interacts with a 50 MeV, 1 nC, 4 ps electron bunch. A scintillator detects up to 600 X-ray photons, with an average energy of 17 keV. In the present status of the experiment Thomson photons have been observed, but the signal to noise ratio is however still too low for an accurate profile measurement.

## 1 INTRODUCTION

A Laser Wire Scanner (LWS) is considered as the most promising option for beam profile measurements at CLIC [1]. Interceptive beam profile monitors, such as Optical Transition Radiation (OTR) screens or solid wire scanners cannot stand high beam current densities without being damaged [2]. Therefore non-degradable diagnostics, like LWS, must be foreseen for both the CLIC Main Beam and Drive Beam. LWS can be very accurate since the resolution is limited by the laser spot size and can be reduced to a few wavelengths. The few microns of transverse size of the CLIC Main Beam can be measured using a UV laser.

LWS are based on the well-known Thomson-Compton scattering [3], where the photons of a laser beam are scattered by incoming electrons. By counting the number of scattered photons as a function of the laser position, the bunch profile can be reconstructed. In a 90° collision the scattered photons spectrum has a maximum at the energy $h\nu_{sc}$, which is given by:

$$h\nu_{sc} = 2\gamma^2 h\nu_0 \qquad (1)$$

where $h\nu_0$ is the laser photon energy and $\gamma$ the relativistic factor of the electrons. Below 1 GeV, the energy of the scattered photons remains small compared to the initial electron energy (Thomson regime). Above this limit, the electron recoil is no longer negligible (Compton regime) and at very high energies the scattered photons take most of the energy of the incident electrons. With the Thomson cross-section, $\sigma_T$, equal to 6.65 $10^{-29}$ m$^2$, very powerful lasers are required to scatter enough photons to allow an accurate detection.

Signal background comes mainly from bremsstrahlung photons created by beam losses. The detection of

scattered photons is therefore much more difficult in the Thomson regime where the scattered photons have energies significantly lower than those of the bremsstrahlung photons.

Only few LWS have been successfully operated around the world so far. At SLAC, on a 50 GeV electron beam, bunch sizes of a few microns have been measured [4]. At Berkeley [5], Thomson photons have been detected using the 50 MeV electron beam and a terawatt Titanium: Sapphire laser. Emittance measurements have been done at the Amsterdam pulse stretcher ring on a 900 MeV electron beam [6]. Recently, at KEK, a laser wire scanner has been developed in order to measure the very small emittances of the 1.28 GeV damping ring [7]. The following sections will describe the LWS tests carried out at CTF II [8] so far.

## 2 EXPERIMENTAL SET-UP

Figure 1 shows a schematic of the experimental layout. A single laser pulse (1047 nm, 4 ps) from a mode locked Nd: YLF laser [8], is split in two parts. One part with 5% of the initial infrared energy is converted into UV by two consecutive doubling crystals. The UV pulse is then directed onto the photo-cathode and produces the electron bunch. The remaining part of the IR beam (95%) is used in the LWS (2.5 mJ). The CTF II photo-injector laser has been especially rearranged (no Drive beam) in this way in order to deliver the maximum laser energy to the LWS. In this set-up, both the electron and laser pulses are in synchronism and the relative timing between the two can be adjusted using an optical delay line. At the gun exit, the electrons enter a 3 GHz-accelerating cavity, which increases their energy up to 50 MeV. They are then focused, using a quadrupole triplet, in the interaction chamber, where the collisions with the laser photons occur. The IR laser beam is focused using a 150 mm focal length lens. The electrons are then deflected by 90° using a dipole magnet. The scattered photons propagate in the forward direction, pass through a 100 μm thick, aluminium window, and are detected using a lead loaded plastic scintillator coupled to a photo-multiplier tube (figure 1b). A considerable amount of lead shielding is placed all around the detector in order to eliminate background coming from the nearby beam dump. The detector was calibrated at the ESRF on the SNBL X-ray line (10-40 keV) [9]. The electron beam current and position are monitored using a pick-up located just before the interaction zone. A bunch charge of typically 1 nC was measured during the tests.

---

*Corresponding author: thibaut.lefevre@cern.ch

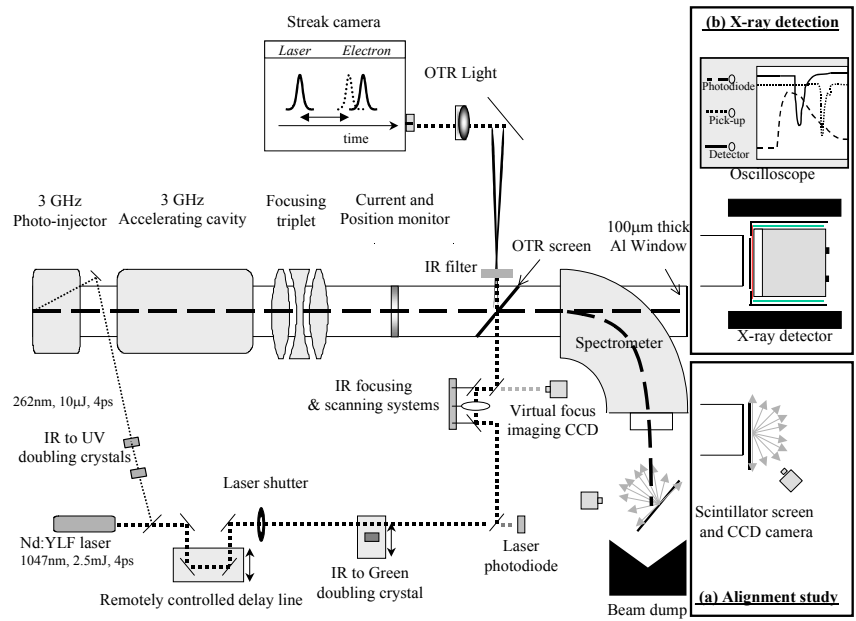

Figure 1: Laser Wire Scanner experimental layout on CTF2

The interaction chamber is also equipped with an aluminium OTR radiator. This is used to measure the electron beam profile ($\sigma_{x,y}$ = 160 µm) and bunch length ($\sigma_z$ =4 ps) by using a streak camera in either focus or sweep mode. The laser energy is monitored using a photodiode detecting the photons leaking through a mirror. The position and the size of the laser spot is also continuously monitored using a CCD camera looking at the virtual focus behind the last mirror. The laser focusing system produces a 30 µm r.m.s spot size with a 5 mm Rayleigh range. It is mounted on a remotely controlled translation stage allowing vertical scanning with steps as small as one micron.

The scattered photons are emitted in a small cone centred in the direction of propagation of the electrons. The detection angle is 26 mrad and according to simulations small angular misalignments, of the order of 5 mrad, can be tolerated. At the beginning of the test (figure 1a), a scintillating screen and a CCD camera were installed in place of the X-ray detector in order to optimise the beam transport to ensure a good alignment.

## 3 OVERLAP TECHNIQUE

Before starting the measurements, the positioning of the laser beam with respect to the electron beam must be adjusted. The streak camera is used to verify the spatial and temporal overlap. For this purpose, the IR pulse is converted into green light and a hole (1 mm diameter) drilled in the centre of the OTR screen allowing the laser light to pass through. In this way both beams can be observed on the same streak camera image. The temporal overlap is achieved by adjusting the LWS laser path using a mirror based delay line, installed on a remotely controlled translation stage. Figure 2a shows a picture obtained with the streak camera in focus mode (2D) and Figure 2b the corresponding image obtained with a sweep

speed of 10 ps/mm. The time interval between the two beams is adjusted to 45 ps, corresponding to the delay introduced by the doubling crystal. We estimate an accuracy of ±3 ps for the time overlap and ±300 µm for the spatial overlap.

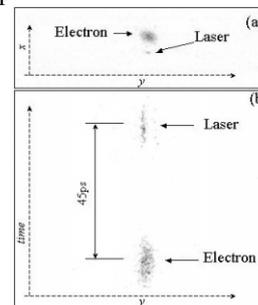

Figure 2: Streak camera images showing the temporal overlap

The doubling crystal is mounted on a remotely controlled translation stage, which allows the overlap of the two beams to be checked without accessing the machine. Small variations in the UV laser position on the photo-cathode or small drifts in the Klystron RF phase have been observed, both leading to changes in the electron timing of a few picoseconds.

## 4 DATA ANALYSIS AND RESULTS

Assuming perfect alignment and the overlap of the two beams, simulations show that 600 photons per machine pulse with an average energy of 17 keV hit the detector. Using the calibration coefficients of our detector we estimate a signal of 3.8 mV. Scans over ±18ps in time or/and ±500 µm vertically, have been performed using steps of 3 ps and 5 µm respectively. For each point of the scan the peak-to-peak values of the X-ray detector, the bunch charge and the laser power are stored.

## 4.1 Background studies

Due to the large background signal observed, data are acquired consecutively with the laser on (30 seconds) and the laser off (10 seconds) for each position of the scan. Laser - off values are used for the background subtraction technique. Figure 3 shows the detector voltage versus bunch charge for three different scans. The slope of the fit line indicates the background level. Using the expected value for the Thomson photons signal (3.8 mV), the signal to noise ratio can be calculated. Large variations from 1/8 to 1/30 have been observed. Above 1 nC, beam losses in the accelerating cavity increase rapidly with direct consequences on the background (dot curve).

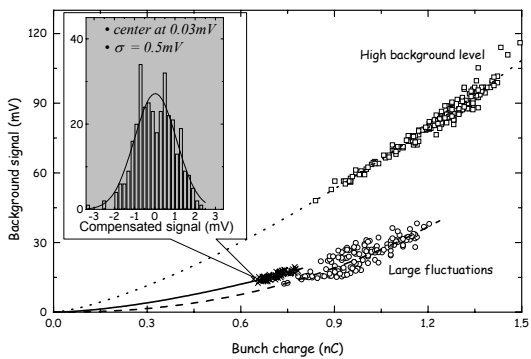

Figure 3: Background signal as a function of bunch charge

In the same figure the histogram of the residuals of the fit is shown. Typical values for the r.m.s of these histograms are between 0.5 and 3.5mV. The background is very sensitive to the position of the beam in the accelerating cavity. Small changes in the position can lead to significant variations in the signal, with no visible effects on the bunch charge.

## 4.2 Results

In Figure 4 the detector signal is plotted as function of time along the scan.

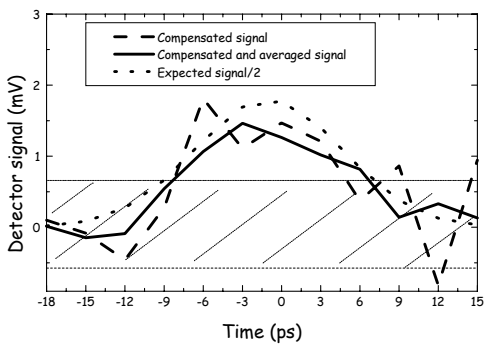

Figure 4: Comparison between measured, smoothed and expected signals for a ±18ps scan

Dash line curve represents the compensated signal and solid line curves the smoothed compensated signal. The two horizontal lines represent the r.m.s amplitude of the random fluctuations of the background signal. The expected signal (dot line) is calculated from the measured parameters. Small offsets in time and position are also introduced in the calculation to fit the measured signal. Maximum values of 2 ps and 175 μm offsets have been observed, which is in good accordance with the precision of our alignment technique. As one can see the X-ray signal is considerably smaller than the expected 3.8 mV mentioned before. The reason for this is not yet clear.

## 5 CONCLUSIONS

A signal correlated with the Thomson photons has been observed. The signal to noise ratio is however still too small to measure a beam profile with sufficient accuracy.

Increasing the measurement time, in order to reduce the statistical error, is unfortunately hindered by the unavoidable fluctuations of the machine parameters.

Background levels are a very important aspect for the design of a LWS. Collimation and background suppression must be carefully investigated, especially for low energy beams. Laser powers higher than 1GW, would be required to obtain a sufficient signal to noise ratio.

## 6 ACKNOWLEDGMENT

The authors would like to thank S. Burger and the Swiss Norwegian Beam line team at ESRF for their help in the calibration of the X-ray detector.

# BEAM PROFILE MEASUREMENTS AT PETRA WITH THE LASERWIRE COMPTON SCATTERING MONITOR

K. Balewski[a], G.A. Blair[b], S. T. Boogert[d], G. Boorman[b], J. Carter[b], T. Kamps[c*],
T. Lefevre[e] H. Lewin[a], F. Poirier[b], S. Schreiber[a] K. Wittenburg[a]

[a] Deutsches Elektron–Synchrotron DESY, D-22603 Hamburg, Germany

[b] Royal Holloway University of London, Egham, Surrey, TW20 0EX, UK

[c] Berliner Elektronenspeicherring-Gesellschaft für Synchrotronstrahlung BESSY, D-12489 Berlin, Germany

[d] University College London, London, WC1E 6BT, UK

[e] CERN, CH-1211 Geneve 23, Switzerland

## Abstract

The vertical beam profile at the PETRA positron storage ring has been measured using a laserwire scanner. A laserwire monitor is a device which can measure high brilliance beam profiles by scanning a finely focused laser beam non-invasively across the charged particle beam. Evaluation of the Compton scattered photon flux as a function of the laser beam position yields the transverse beam profile. The aim of the experiment at PETRA is to obtain the profile of the positron beam at several GeV energy and several nC bunch charge. Key elements of laserwire systems are currently being studied and are described in this paper such as laser beam optics, a fast scanning system and a photon calorimeter. Results are presented from positron beam profile scans using orbit bumps and a fast scanning scheme.

## INTRODUCTION

Future high performance TeV-scale lepton collider as well as high brilliance linac based light sources require on-line, non-invasive beam size monitors with micron and sub-micron resolution for beam phase space optimization [1]. A laserwire monitor is a device where a finely focused, high power laser beam is scanned transversely over the lepton beam. The resulting Compton-scattered photons are detected downstream and the measurement of the total energy of these photons as a function of laser spot position yields the lepton bunch transverse dimensions [2]. Laserwire beam profile monitors have been tested at the SLC at SLAC [3] and ATF at KEK [4]. The aim of the experiment at PETRA is to elevate these designs and to investigate key issues for a laserwire device in order to develop a standard diagnostic tool for low-disruption, high-resolution beam profile measurements.

## EXPERIMENTAL SETUP

The laserwire experiment is installed at the storage ring PETRA at DESY. The PETRA ring operates as pre-accelerator for positrons and protons and serves the collider HERA. In the context of an upgrade to a 3rd genera-

tion synchrotron light source some beam time is allocated to machine development. PETRA was chosen for experiments with the laserwire because of the availability of a long straight section for hardware installation, an existing access pipe, sufficient energy and because of its bunch pattern, which is similar to high energy linear collider. Beam tests with the laserwire were carried out at 7 GeV with average bunch currents of 7.1 mA and 40.5 mA. The laser pulses were triggered to interact with the first bunch of the bunch train carrying bunch charges of 3.9 nC and 22.3 nC. From the optics lattice the average beam size in the ring is $\sigma_x = 268$ $\mu$m for the horizontal and $\sigma_y = 68$ $\mu$m for the vertical dimension.

The experimental setup is sketched in Fig. 1. The setup is

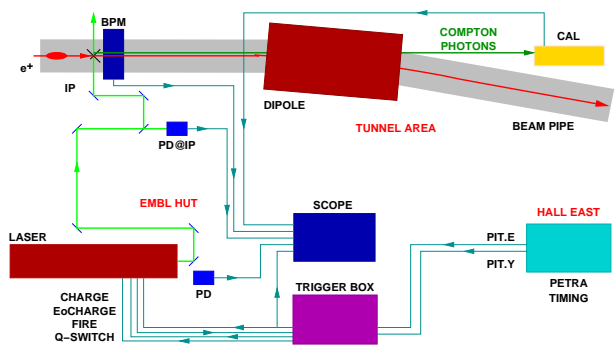

Figure 1: Positron beam, laser and trigger path for the laserwire experimental setup.

mainly divided in two areas, the PETRA accelerator tunnel and the laser hut. The trigger signals for synchronization of the laser and positron beams are derived from the PETRA accelerator bunch clock and brought to the laser hut from the access hall. These are then fed into VME-based trigger electronics, which exchanges trigger and status signals with the laser. The laser timing is measured relative to a BPM signal at the IP using two fast photodiodes; one in the laser hut and the other close to the IP. The Compton-scattered photons are boosted along the incoming positron beam direction and exit through the beampipe wall within a downstream dipole magnet. Most of the photons interact with the wall material producing an electromagnetic

shower. The tail end of the shower is measured by a lead-tungstate calorimeter read out by a fast ADC.

### Laser Beam

The laser pulses are created in a Q-switched Nd:YAG laser amplifier with second harmonic generator. The average power is 8 W in the IR at a wavelength of 1064 nm and 2.1 W in the green at 532 nm. Green light was chosen for this experiment because its shorter wavelength enables both smaller spot-sizes to be achieved and greater energy deposits in the calorimeter. The longitudinal profile was measured using a streak camera with 5 ps time resolution. The data revealed a pulse length of $\Delta t = 12.5$ ns FWHH with a sub-structure of roughly 70 ps peak-to-peak and 70 ps peak width at full contrast. This sub-structure is due to mode-beating of different longitudinal modes lasing and causes the Compton signal amplitude to vary between zero and full signal for different laser shots. The transverse profile of the laser beam was measured in the near and far-field with knife edge and sliding slot techniques. For the far-field measurement the beam was focused using a $f = 125$ mm doublet focusing lens of the same type as used for the IP focusing. A viewport window was also included in the setup. The mode quality parameter was measured to be $M_y^2 = 8.5 \pm 0.6$ for the vertical and $M_x^2 = 5.6 \pm 0.4$ for the horizontal dimension. The measured laser waist radius (at $1/e^2$ intensity) is $w_y = 77 \pm 5$ $\mu$m and $w_x = 69 \pm 6$ $\mu$m.

The laser pulses are transported via a matched Gaussian relay made up of two $f = 5$ m lenses over a distance of 20 m from the laser hut via an access pipe into the tunnel housing the accelerator. The laser beam passes then the scanning mirror before it reaches a focusing lens with $f = 125$ mm back-focal length. The scanner is a piezo-driven platform with an attached 25 mm high-reflectivity mirror. The maximum loaded frequency of the platform is 1 kHz with a scan range of $\pm 2.5$ mrad. The optical elements before the laser-positron IP are shown in Fig. 2. After the interaction the main part of the beam intensity is divided and guided into an appropriate dump. The remaining intensity is used for diagnostics and relayed on a CCD camera for online monitoring of the laser spot size and position at the IP.

### Compton Calorimeter

The Compton photon calorimeter is composed of lead-tungstate (PbWO4) crystals fixed with optical grease to a matching square face photomultiplier. The individual crystals have dimensions of $18 \times 18 \times 150$ mm and are arranged in a 3 by 3 matrix (see [5] for more details). From calibration measurements with a testbeam from the DESY II accelerator, the complete detector setup including ADC readout was tested with electrons from 450 MeV to 6 GeV. The energy resolution was found to be better than 6% for individual crystals and 10% for the overall setup. Simulations show that with the 3 by 3 matrix 95% of the total energy de-

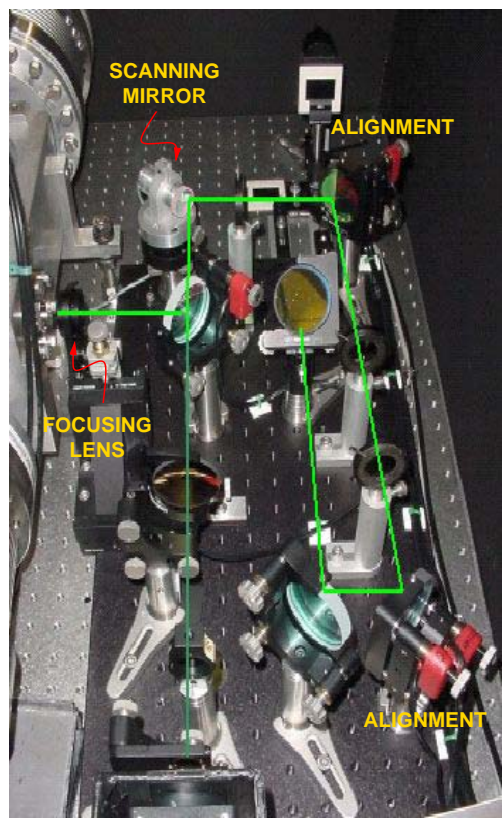

Figure 2: Optical elements and laser pathway before the interaction point with the positron beam.

posit is collected for an incoming Compton-scattered photon with 300 MeV energy.

## DATA TAKING

To establish initial overlap between positron and laser beam first the timing was fixed by fine-adjusting the laser pulse timing with a delay box relative to the BPM signal. Then a local orbit bump was driven to bring both beams to transverse overlap. This procedure usually takes less than 5 min. After that the laser beam was scanned using the piezo-driven platform over the positron beam. At each scan point 5000 events were recorded before the beam was moved again. In Fig. 3 a typical background and signal spectra are displayed. The low energy pedestal in both spectra corresponds to the electronics base. With a laser repetition rate of 30 Hz one scan point is handled in 3 min resulting in 30 min to complete a full scan. During the scan the orbit stability is observed with the BPM at the interaction point. The orbit was found to be stable within 40 $\mu$m.

## DATA ANALYSIS

After applying a background cut eliminating the synchrotron radiation pedestal the individual spectra at the different scan points were integrated. In Fig. 4 the resulting total energy deposit versus laser beam position is plot-

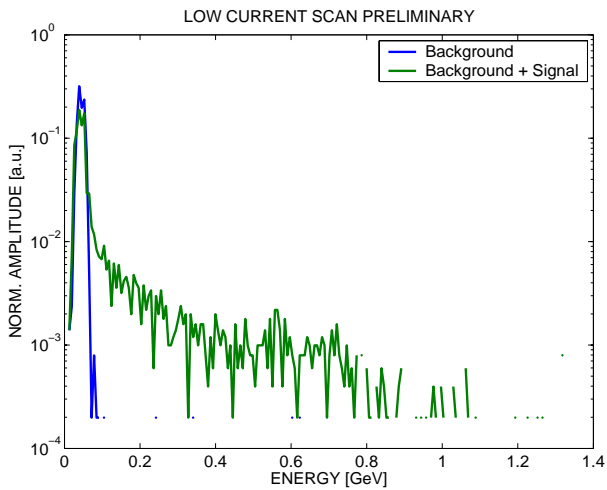

Figure 3: Preliminary data for background and signal events from the low current scan.

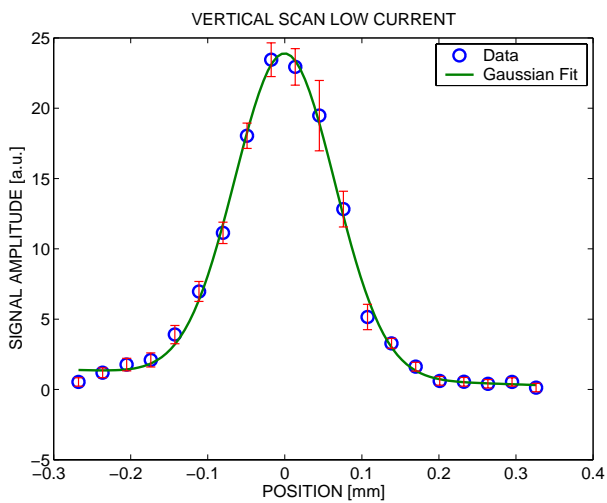

Figure 4: Spotsize measurement data and fit for the low current scan.

ted for the low current scan. For the high current scan a similar results was achieved. This distribution can be approximated with a Gaussian function with a linear gradient term. The linear term in the model compensates for the decreasing beam current over the 30 min measurement time. The measured beam size for the low current scan is $\sigma_m = 68 \pm 3_{stat} \pm 14_{sys}$ $\mu$m and for the high current scan $\sigma_m = 80 \pm 6_{stat} \pm 16_{sys}$ $\mu$m. The systematic error is dominated by the uncertainty of the set-voltage to angle scanner calibration and will be investigated with dedicated measurements.

## COMPARISON WITH SIMULATIONS

The propagation of the Compton photons was simulated using the Geant4 [6] simulation package including background sources like synchrotron radiation and bremsstrahlung. The accelerator environment is modeled including beampipe, magnets, and cooling water channel. From an analytical description of the Compton process [7] the number of Compton photons can be estimated for the relevant laser and positron beam parameters. For the low current case with 3.9 nC in total 98 photons are released and the Compton edge is at 1.17 GeV resulting in 57 GeV total energy deposit, for the high current case with 22.3 nC in total 561 photons are released resulting in 328 GeV energy deposit. Most of the energy is lost in the beampipe material as the Compton photons hit the chamber under a shallow angle inside the dipole magnet with an effective length of roughly 100 cm. In the next step the measured longitudinal profile of the laser beam will be included in the simulations.

## CONCLUSION AND OUTLOOK

A laserwire monitor system has been setup and run at the PETRA accelerator. The vertical beam size of the positron beam was measured using a scanning platform, results agree well with expectations from lattice calculations. The laser has been upgraded with a longitudinal mode filtering etalon. For the Compton photons pathway planning and construction is under way to replace the dipole vacuum chamber with a chamber with a thin exit window, to enhance the observed signal.

# BEAM PROFILE MEASUREMENTS AND SIMULATIONS OF THE PETRA LASER-WIRE

J. Carter, I. Agapov, G. A. Blair, G. Boorman, C. Driouichi, F. Poirier
M. T. Price (Royal Holloway University of London, Surrey), T. Kamps (BESSY GmbH, Berlin),
K. Balewski, H. Lewin, S. Schreiber, K. Wittenburg (DESY, Hamburg),
N. Delerue, D. F Howell (University of Oxford, Oxford), S. T. Boogert, S. Malton (UCL, London) *

## Abstract

The Laser-wire will be an essential diagnostic tool at the International Linear Collider. It uses a finely focussed laser beam to measure the transverse profile of electron bunches by detecting the Compton-scattered photons (or degraded electrons) downstream of where the laser beam intersects the electron beam. Such a system has been installed at the PETRA storage ring at DESY, which uses a piezo-driven mirror to scan the laser-light across the electron beam. Latest results of experimental data taking are presented and compared to detailed simulations using the Geant4 based program BDSIM.

## INTRODUCTION

The International Linear Collider (ILC) will be a TeV-scale lepton collider that will require non-invasive beam size monitors with micron and sub-micron resolution for beam phase space optimisation [1]. Laser-wire monitors operate by focussing a laser to a small spot size that can be scanned across the electron beam, producing Compton-scattered photons (and degraded electrons). These photons can then be detected further downstream using the total energy observed as a function of laser spot position to infer the transverse profile of the electron bunch. The Laser-wire system installed in the PETRA ring is part of an ongoing effort in the R&D of producing a feasible non-invasive beam size diagnostic tool.

## EXPERIMENTAL SETUP

The PETRA accelerator was chosen for the installation of the Laser-wire experiment because it is capable of producing bunch patterns similar to the ILC. Laser-wire tests are run using a 7 GeV positron beam with a single bunch with a charge of typically 7.7 nC. From the optics lattice the average beam size is $\sigma_x = 268$ $\mu$m for the horizontal and $\sigma_y = 68$ $\mu$m for the vertical dimension.

Preliminary simulations showed that the Compton-scattered photons loose the majority of their energy in the material of the dipole magnet's beampipe due to hitting the wall with a shallow angle, resulting in an effective length

of 100 cm of Aluminium. An exit window was therefore designed and installed (by DESY) to allow these photons to reach the detector with little deterioration (see Fig. 1).

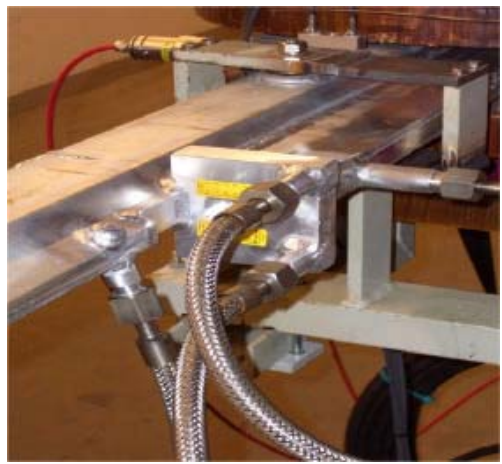

Figure 1: New exit window for Compton photons

### Laser Beam

The laser pulses are created in a Q-switched Nd:YAG laser operating at 532 nm. The pulses are then transported via a matched Gaussian relay made up of two lenses over a distance of 20 m from the laser hut via an access pipe into the tunnel housing the accelerator. The laser beam is then reflected off the scanning mirror before it reaches a focusing lens with $f = 117$ mm back-focal length. The scanner is a piezo-driven platform with an attached high-reflectivity mirror which has a maximum scan range of $\pm 2.5$ mrad. The peak power at the laser exit was measured to be 3.63 MW. At the IP the peak power is reduced to 1.46 MW as higher order modes carry some fraction of the beam power but these are focussed out of beam transport, which is only matched for the fundamental mode. The longitudinal profile was measured using a streak camera with 5 ps time resolution. The data revealed a pulse length of $\Delta t = 12.5$ ns FWHM with a sub-structure of roughly 70 ps peak-to-peak and 70 ps peak width at full contrast due to mode-beating. This causes the Compton signal amplitude to vary between zero and full signal for different laser shots. In order to reduce the data taking time the current laser will be replaced with an injection seeded system enabling faster data taking.

## Compton Calorimeter

The Laser-wire set up makes use of a calorimeter composed of 9 Lead Tungstate ($PbWO_4$) crystals arranged in a $3 \times 3$ matrix fixed with optical grease to a square faced photomultiplier. The individual crystals have dimensions of $18 \times 18 \times 150$ mm. The complete detector set up was tested with a testbeam from the DESY II accelerator using electrons from 450 MeV to 6 GeV. Energy resolution was shown to be better than 6% for individual crystals and 10% for the overall set up. Simulations show that for the $3 \times 3$ matrix, 95% of the total energy deposit is collected for an incoming Compton-scattered photon with 300 MeV energy [2].

## Data Acquisition

The Laser-wire DAQ system has two main components: the hardware trigger which synchronises the laser and DAQ components to the electron (positron) bunch, and the software which controls the acquisition and collation of data from each sub-component of the system.

The hardware trigger operates with two inputs from the PETRA Integrated Timing system (PIT) and produces the necessary signals to fire the laser. The trigger card also produces a signal to trigger the CCD cameras and a signal to start the software data acquisition. When the signal from the trigger card is received a counter which runs for approximately 420 $\mu$s is started. After this time a signal is sent to the integrator card, lasting around 50 $\mu$s, to integrate the output from the calorimeter. The integrated signal is read by an ADC channel.

The DAQ software also produces a programmable signal, up to a peak of 10 V, which is amplified by a factor of 10 and this is used to drive the piezo-electric scanner. A scaled version of the scanner amplifier output is read by an ADC channel. The other sub-components of the DAQ system: the BPM monitor, the PETRA data monitor and the CCD cameras are also read out. Communication with each component is performed by a messaging system using TCP/IP.

# DATA ANALYSIS

## Laser Beam Size

In order to determine the transverse size of the electron beam, it is necessary to know the properties of the laser that is being used to scan. Particular attention is paid to the spot size at the laser waist, $\sigma_0$, and the Rayleigh range, $z_R$, (the distance from the waist at which the beam size $\sigma = \sqrt{2}\sigma_0$). These properties are related by Eq. 1:

$$\sigma = \sigma_0 \sqrt{1 + \left(\frac{z}{z_R}\right)^2} \qquad (1)$$

where $z_R = \frac{4\pi\sigma_0^2}{M^2\lambda}$.

The laser is focused using the same final focus lens as described previously. A CMOS camera is placed on a track rail so that it can be moved through the focal plane parallel to the beam direction. Due to the high power of the laser, the beam was first passed through a 99.9 % reflective mirror, and then through a variable amount of neutral density filter in order to prevent saturation and damage to the camera pixels. The camera was moved along the track rail to a number of positions, and 100 images were taken in each location.

The images taken by the camera are stored as 8-bit greyscale bitmap files. The pixel data is projected onto the y-axis, and fitted to a gaussian on a linear background in the region around the signal peak. The width at each location is then plotted, and fitted to Eq. 1. From this we obtain $M^2 = 7.6 \pm 0.41$, which is within the expected range, and $\sigma_0 = (35 \pm 2)$ μm, as shown in Fig. 2.

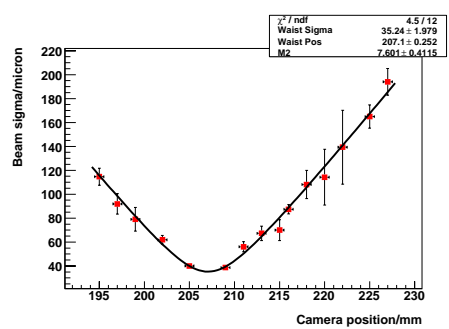

Figure 2: Variation in transverse beam size of the laser around the focus at the IP.

## Scan Data

The laser is scanned across the electron beam by tilting a mirror on a piezo-electric stack to produce a deflection of $\pm 2.5$ mrad. Focusing through the lens produces a travel range for the focal spot at the IP of 585 $\mu$m. The scanner voltage is applied in a stepped sinusoidal pattern; 10 triggers are taken at each of 100 voltages over a whole $2\pi$. The trigger signal is taken from the laser trigger card running at 30 Hz, so a full scan takes approximately 33 s.

The signal from the ADC is expected to display two peaks; one as the laser crosses the electron beam on a rising voltage to the scanner, and one on a falling voltage. The trigger number exactly half way between the peaks should correspond to a turning point in the scanner position. The mean of the background subtracted ADC counts at each voltage is then fitted to a gaussian whose width, $\sigma_m$ is given by $\sigma_m^2 = \sigma_e^2 + \sigma_0^2$. Fig. 3 shows the typical results observed for a single scan and the results of multi-scan shifts are presented in Table 1. Note that the large signal variation in Fig. 3a is partly due to the sub-structure of each laser pulse and will be removed by a better laser.

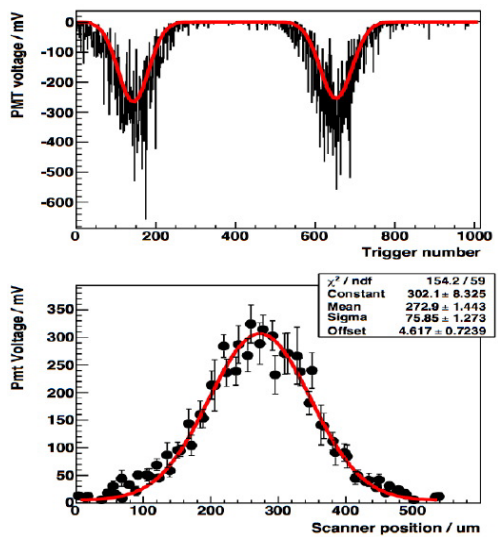

Figure 3: a: PMT voltage vs trigger number, fitted to a constant background with two gaussians. b: Mean PMT voltage vs laser focus position with a fit showing $\sigma_m$.

| Shift | No. of Scans | $\sigma_e \ [\mu m]$ |
|-------|-------------|---------------------|
| 1 | 7 | $62.89 \pm 2.45$ |
| 2 | 7 | $71.67 \pm 3.28$ |
| 3 | 3 | $77.22 \pm 5.51$ |

Table 1: Data run results for extracted electron beam size, $\sigma_e = \sqrt{\sigma_m^2 - \sigma_0^2}$. The errors are the RMS from several scans

## COMPARISON WITH SIMULATIONS

The entire PETRA Laser-wire set up has been simulated using BDSIM [3], which is a fast tracking code utilising the Geant4 [4] physics processes and framework. The simulation is a full model of the accelerator components including beampipe, magnets, and cooling water channel. For each simulated event a Compton scattered photon is generated with an energy based upon the Compton Spectrum predicted for the PETRA Laser-wire parameters. This photon is tracked to the detector whilst fully simulating any interactions with materials such as the beampipe wall. This process is repeated to create an effective single Compton energy distribution and its corresponding distribution at the detector after passing through any matter along the photon path.

The single photon distribution in the detector is extrapolated to the $N_{photon}$ spectrum using Poisson statistics whilst also accounting for the energy resolution of the calorimeter and the longitudinal sub-structure of a typical laser pulse. The simulated spectrum is compared directly to the experimental data (see Fig. 4), where the laser and electron beam were well aligned. The expected number of Compton-scattered photons, $N_{photon}$, per shot with the Laser-wire setup parameters is approximately $170 \pm 25$ photons, which agrees with the theoretical value.

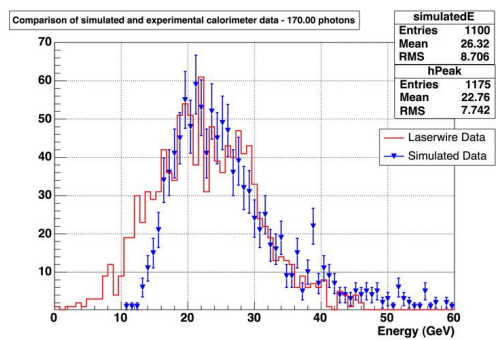

Figure 4: Calorimeter energy spectra for data and simulated events.

The experimental data show an energy resolution of 34% which is dominated by the longitudinal fluctuations in the laser power. The simulation models these fluctuations using relatively old streak camera data as described above and so does not account for degradation in the quality of the laser since then. The calorimeter has also not been calibrated for the range of energy deposits now incident upon it and has been in the PETRA radiation environment for three years. This could explain why the simulation fails to completely model the experimental data in the lower energy region.

## OUTLOOK

The future strategy for the Laser-wire project can be characterised in the short term to concentrate on non-laser issues like data acquisition, signal detection, vertical scanning, and implementation into a linac beamline. This aims at the development of a standard diagnostic tool to be placed at many locations along the accelerator beamline. In the long run R&D work is planned to develop a laser system producing pulses matching the ILC micro pulse structure. Here the target is to have a beamsize monitor with full flexibilty. To meet the short-term targets it is planned to purchase an injection seeded Q-switch laser with second harmonic generation having excellent longitudinal and transverse mode quality. A complimentary project concentrating on the achievement of micron-scale laser spot-sizes is underway at the Accelerator Test Facility (ATF) at KEK.

# BEAM PROFILE MEASUREMENTS WITH THE 2-D LASER-WIRE *

M.Price, G.Blair, A.Bosco, G.Boorman, J.Carter, I.Agapov, C.Driouichi[†],
S.Boogert (John Adams Institute at RHUL, London, UK.),
T.Kamps (BESSY GmbH, Berlin, Germany.),
F. Poirier, K.Balewski, H.Lewin, S.Schreiber, K.Wittenburg (DESY, Hamburg, Germany.),
N.Delerue, D.Howell (University of Oxford, Oxford, UK.)

*Abstract*

The laser-wire will be an essential diagnostic tool at the International Linear Collider (ILC). It uses a finely focused laser beam to measure the transverse profile of electron bunches by detecting the Compton-scattered photons downstream of the interaction point (IP), where the laser beam intersects the electron beam. Such a system has been installed at the PETRA storage ring at DESY, which uses a piezo-driven mirror to scan the laser-light across the electron beam. This paper reports recent upgrades to the PETRA system, including the implementation of a new laser.

## INTRODUCTION

The ILC will be a TeV-scale electron-positron collider that will require non-invasive beam size monitors with micron and sub-micron resolution for beam phase space optimization [1]. Laser-wire monitors operate by focussing a laser to a small spot size that can be scanned across the electron beam, producing Compton-scattered photons (and degraded electrons). The photons can then be detected further downstream using a photon calorimeter (which measures their energy). The total energy observed as a function of the laser spot position is used to infer the transverse profile of the electron bunch. The Laser-wire system installed in the PETRA ring is part of an ongoing effort in the R&D of producing a feasible non-invasive beam-size diagnostic tool.

## EXPERIMENTAL SETUP

PETRA Laser-wire experiments use a 7.2 GeV positron (or electron) beam with a single-bunch with a charge of 7.7 nC. Optics lattice studies suggest that the average beam size is $\sigma_x = 268\ \mu m$ for the horizontal and $\sigma_y = 68\ \mu m$.

The laser-wire experiment must coordinate the arrival of high-energy laser pulses with the arrival of a targeted electron bunch at the IP, and record the resulting calorimeter measurements from the Compton photons resulting from scattering at the IP. This is all achieved using the laser-wire data acquisition system (DAQ). Figure 1 shows a schematic representation of the laser-wire experiment and its signal coordination.

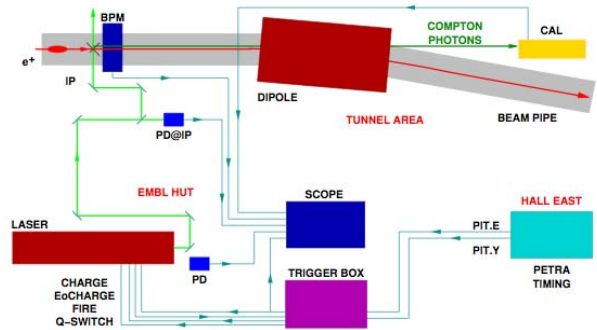

Figure 1: A schematic representation of the electron/positron beam, laser light path and signal coordination for the laser-wire experiment.

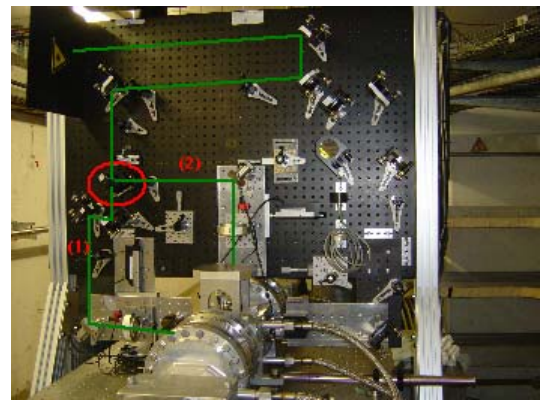

Figure 2: The vertical breadboard arrangement of the Laser-wire experiment: capable of scanning the vertical (path 1) and horizontal (path 2) profiles of the electron beam, at any one time. The pre-IP laser beam path is drawn in green. The mirror flipper, which selects the axis of the electron beam to scan, is circled in red.

An upgrade to the previous 1-dimensional PETRA laser-wire system [3] is reported here. The 2-D scanning *vertical breadboard* (see figure 2) was installed at PETRA in December 2005. Laser pulses arrive at a mirror flipper, which is pre-set to send the pulses along path (1) or (2), scanning the electron's vertical or horizontal profile, respectively. After the mirror flipper the pulses are reflected off a piezo-crystal driven mirror onto a LAP250 lens [1], where they are

† Now at Niels Bohr Institute, Copenhagen.

[1]A two-inch diameter compound lens of 250mm focal length

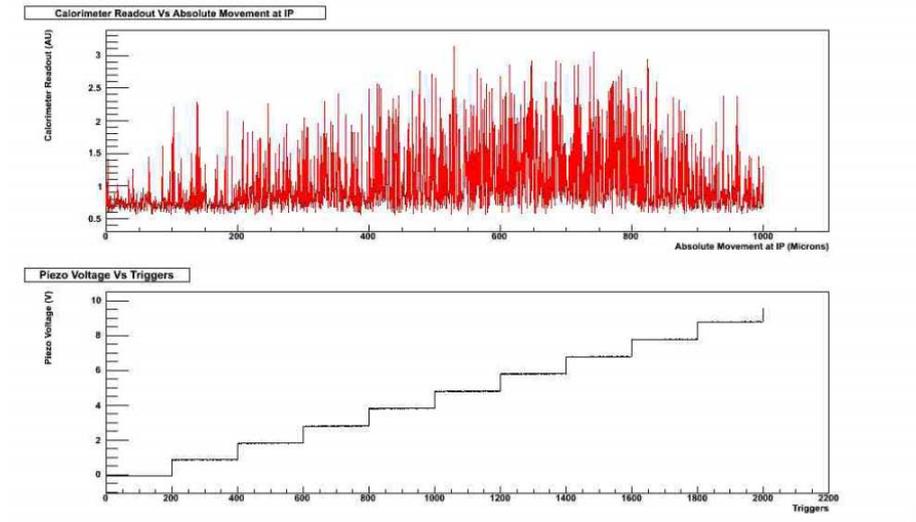

Figure 3: Data taken with the injection seeder turned off. The top plot shows the calorimeter output (arbitrary units) versus the laser spot position at the IP (in microns). The bottom plow is the corresponding piezo scanner voltage (Volts) versus the trigger number (laser shot number).

then focused onto the electron bunch. Changing the voltage across the piezo-crystals changes the angle at which the laser pulses are reflected through the LAP250 lens, thereby controlling the position at which the laser pulses are swept across the electron bunch. By such means, laser pulses are swept across the electron bunch.

The non-interacting photons continue onto the post IP section of the breadboard. Here the remaining laser pulse energy is measured using an energy meter, and the cross-sectional profile of the laser pulse is measured using a CCD camera. This information is used to monitor laser quality.

### Laser

Laser pulses of $\lambda = 532\ nm$ are generated by a newly acquisitioned injection-seeded, Q-Switched Nd:YAG laser firing into a second harmonic separating box[2]. The injection seeding is desired to eliminate mode-beating effects from the laser pulses, producing a more uniform photon intensity in the pulse. A summary of the laser characteristics are shown in table (1).

Initial testing showed the laser exhibited poor modal quality, directional instability, and significant pulse-to-pulse time jitter and power fluctuation. These effects have since been attributed to a damaged Brewster plate and spurious reflections off its internal (single mode) pinhole. Further testing is currently underway to correct and permanently fix the problems.

### Compton Photon Detector

The Compton photon detector is composed of a 3 x 3 matrix of lead tungstate ($PbWO_4$) crystals, fixed to a matching face photomultiplier. Each crystal has dimensions of

---

Table 1: The table summarizes the results of initial testing on the new laser.

| Summary of laser Measurements | |
|---|---|
| Laser pulse freq. | 20 Hz |
| $M^2$ | 1.4 |
| Pulse duration | 5 ns |
| Pulse jitter (trig. Sync_out) | 2-6 ns |
| Pulse jitter (trig. PIT) | 2-6 ns |
| Energy (532nm, pulse-to-pulse) | 3.5 mJ $\pm$ 0.875 mJ |
| *Equiv. Power (pulse-to-pulse)* | 0.18 W $\pm$ 0.044 W |

18 x 18 x150 mm and an energy resolution of approximately 5.4% measuring electrons at 6 GeV [2]. Measuring the same electrons, the total system had an energy resolution of approximately 10%. Simulations show that Compton-scattered photons (300 MeV) deposit their energy with 95% efficiency on the 3 x 3 crystal matrix.

## DATA-TAKING

The laser spot was scanned across the electron bunch in 11 steps[3], there were 2048 laser shots, and hence, 2048 triggers or data points per step. Each step corresponded to approximately 46 microns at the IP.

Large time jitters would cause the laser pulses to miss the electron bunch entirely and therefore result in no Compton photons. However a small enough time jitter could result in some overlap between the electron bunch and laser pulse, and hence Compton photons. Such fluctuations can be observed in figures 3 & 4.

---

[2]Surelite Separation Package (SSP).

[3]Each step corresponds to 1V of the piezo scanner, which has a full scale deflection range of 0-10V.

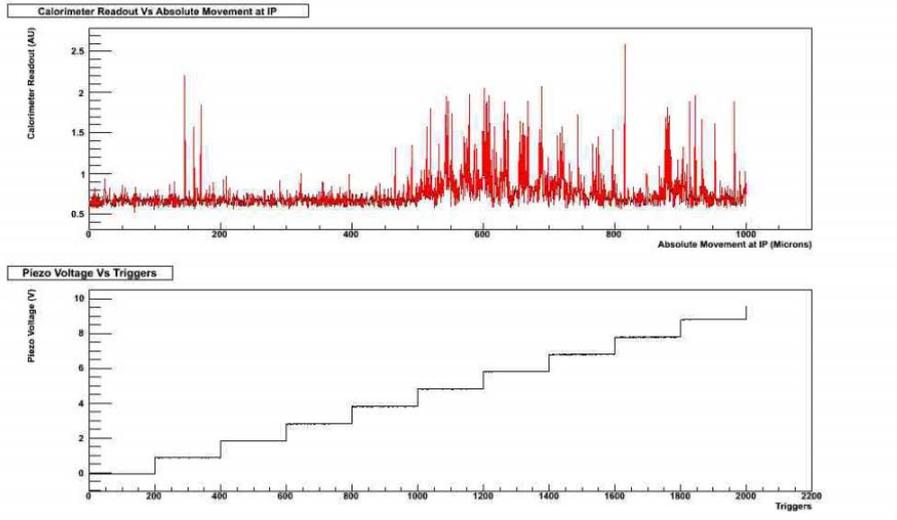

Figure 4: Data taken with injection seeding turned-on. The top plot shows the calorimeter output (arbitrary units) versus the laser spot position at the IP (in microns). The bottom plot shows the corresponding piezo scanner voltage (Volts) versus the trigger number (laser shot number).

## DATA ANALYSIS

Figures 5 & 6 show the data analysis for the unseeded and seeded experiments, respectively. The plots are fit with a compound function consisting of: *a Gaussian*[4] and *a polynomial function of order 1 (straight line with a gradient and y-intercept)* [5]

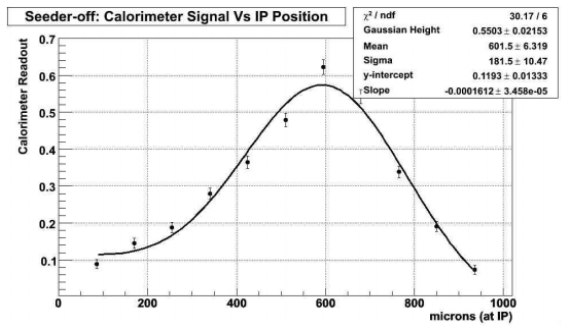

Figure 5: Data analysis of the experiment with no injection seeding. The plot shows 11 points, where each point is the averaged calorimeter readout per piezo voltage, versus the movement at the IP (in microns).

## CONCLUSION & OUTLOOK

Compton interactions have been produced using the 2-D laser-wire breadboard in conjunction with the new Nd:YAG laser. The results with the new system agree closely with those expected for the predicted electron bunch dimensions

---

[4]Due to the convolution of a Gaussian laser pulse with a Gaussian electron bunch [4]

[5]A short-time approximation to the exponential decay of the electron beam with time.

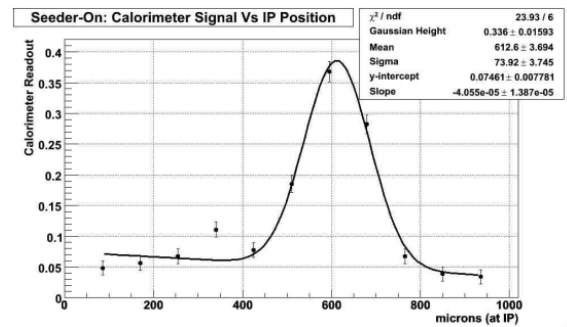

Figure 6: Data analysis of the experiment with injection seeded turned-on. The plot shows shows 11 points, where each point is the averaged calorimeter readout per piezo voltage, versus the movement at the IP (in microns).

at the IP. Further steps are being taken to reduce the time jitter of the laser to allow a greater scanning efficiency. Studies benchmarking the laser-wire against standard (invasive) carbon wire-scans are currently underway. Research is also being carried out on fast scanning techniques using crystals whose refractive index varies with electric field [5].



\end{document}